\documentclass[prb,aps,twocolumn,superscriptaddress,reprint,showpacs,floatfix,longbibliography]{revtex4-2}
\usepackage{amsmath,amssymb,amsfonts,bbm,float,graphics,epsfig,epstopdf,color,verbatim,tabularx,bm,multirow,hyperref,ulem,times,subfigure}
\hypersetup{
  colorlinks, linkcolor={blue},
  citecolor={blue}, urlcolor={blue}
}

\usepackage{textcomp}
\usepackage{times}
\usepackage{tabularx}
\usepackage{graphicx}
\usepackage{float}
\usepackage{latexsym,amsmath,amssymb,bm,euscript}
\usepackage{color}
\usepackage{subfigure}
\usepackage{epstopdf}

\graphicspath{{./Figs/}}

\begin{document}
\title{Vestigial Gapless Boson Density Wave Emerging between $\nu=1/2$ Fractional Chern Insulator and Finite-Momentum Supersolid}
\author{Hongyu Lu}
\affiliation{Department of Physics and HK Institute of Quantum Science \& Technology, The University of Hong Kong, Pokfulam Road, Hong Kong SAR, China}
\author{Han-Qing Wu}
\affiliation{Guangdong Provincial Key Laboratory of Magnetoelectric Physics and Devices, School of Physics, Sun Yat-sen University, Guangzhou 510275, China }
\author{Bin-Bin Chen}
\affiliation{Department of Physics and HK Institute of Quantum Science \& Technology, The University of Hong Kong, Pokfulam Road, Hong Kong SAR, China}
\author{Zi Yang Meng}
\email{zymeng@hku.hk}
\affiliation{Department of Physics and HK Institute of Quantum Science \& Technology, The University of Hong Kong, Pokfulam Road, Hong Kong SAR, China}

\begin{abstract}
The roton-triggered charge-density-wave (CDW) is widely studied in fractional quantum Hall (FQH) and fractional Chern insulator (FCI) systems, and there also exist field theoretical and numerical realizations of continuous transition from FCI to superfluid (SF). 
However, the theory and numerical explorations of the transition between FCI and supersolid (SS) featuring coexistence of CDW and SF orders, are still lacking.
In this work, we study the topological flat-band lattice models with $\nu=1/2$ hard-core bosons, where previous studies have discovered the existence of FCI states and possible direct FCI-SS transitions~\cite{DNSheng2011boson,Luo2020boson}.
While the FCI is robust, we find the direct FCI-SS transition is absent, and there exist more intriguing scenarios.
In the case of checkerboard lattice, we find an intermediate {\it gapless} CDW state without SF, sandwiched between FCI and SS. 
This novel state is triggered by the roton instability in FCI and it further continuously brings about the intertwined finite-momentum SF fluctuation when the CDW order is strong enough, eventually transiting into an unconventional {\it finite-momentum} SS state that is similar to pair-density-wave superconductors. 
The intermediate gapless CDW state is a {\it vestige} from the SS state, since the increasing quantum fluctuation melts only the Larkin-Ovchinnikov-type SF order in SS but its (secondary) product -- the CDW order -- survives.
On honeycomb lattice, we find no evidence of SS, but discover an interesting sequence of FCI-Solid I-Solid II transitions, with both solids incompressible. 
Moreover, in contrast to previous single-roton condensation, this sequence of FCI-Solid I-Solid II transitions is triggered by the softening of multi-roton modes in FCI. The minimal roton mode in FCI gives rise to Solid I, and when this CDW is strong enough, a higher-energy mode in FCI goes soft and renders the doubled translation period of Solid II.
Considering the intertwined wave vectors of the CDW orders, Solid I is also a {\it vestige} of Solid II.
Our work provides new horizons not only for the quantum phase transitions in FCI but also for the intertwined orders and gapless states in bosonic systems, which will inspire future theoretical and experimental studies.
\end{abstract}

\date{\today }
\maketitle

% ================================ %
\section{Introduction}
In bosonic systems, the properties of fractional quantum Hall (FQH) states and fractional Chern insulators (FCI) are well studied~\cite{Stormer1999FQH, Klaus2020FQH, Laughlin1983FQH, Haldane1983FQH, GMP1985, Regnault2004boson, Jain2007composite, Cooper2009composite, DNSheng2011boson, Luo2020boson, Wang2012nonabelian, Repellin2014SMA, Zhu2016FQAH, Zeng2022FQAH}, and the $\nu=1/2$ FQH states are experimentally realized in ultracold atom~\cite{Julian2023photonFQH} and quantum electrodynamics~\cite{Wang2024photonFQH} lattice systems. 
More broadly, the quantum phase transitions related to these topologically ordered states  
 have attracted widespread attention in the past few decades, with ongoing research continuously extending our knowledge of the interplay between FQH/FCI and symmetry-breaking states. 
 In both fermionic and bosonic FQH/FCI states, the softening of magneto-roton modes at finite momentum can lead to phase transitions to charge density waves (CDW)~\cite{GMP1986,  Kumar2022_condense, LHY2024_roton_transition, LHY2024_FQAHS_transition} and such CDW orders could even coexist with the topological order without closing the charge gap or changing the fractional Hall conductivity~\cite{LHY2024_FQAHS_transition}.
 In bosonic systems, another exotic case is the continuous FQH/FCI - superfluid (SF) transition~\cite{Barkeshli2012CFL_FL, Barkeshli2014FQH-SF,Barkeshli2015FCI-SF, Song2023deconfined, songPhase2024}, whose theoretical interpretation is believed to be beyond the Landau-Ginzburg paradigm. The field theories have been proposed with predicted emergent symmetry and fluctuations~\cite{Song2023deconfined, songPhase2024}, and this continuous transition has been numerically realized recently~\cite{LHY2024FCI_SF}.

However, despite these progresses (FCI-SF transitions and fermionic FCI-CDW transitions driven by the magnetoroton mode), the mechanism of transition from FQH/FCI to supersolid (SS) state with coexisting SF and CDW, is still unclear and it is unique in bosonic systems with much less explorations compared to the fermionic FCI systems.
One could, in principle, based on simple analysis propose three different scenarios, which are intuitive and interesting regardless of the order of transitions.
The straightforward scenario is a direct FCI-SS transition, where the CDW and SF orders form simultaneously. 
To the best of our knowledge, there exist no effective theories or clear numerical demonstrations of such a direct transition between the FCI and SS states. 
The second scenario is a FCI-SF-SS sequence. Considering the successful realization of continuous FCI-SF transition~\cite{Barkeshli2012CFL_FL, Barkeshli2014FQH-SF,Barkeshli2015FCI-SF, Song2023deconfined, songPhase2024,LHY2024FCI_SF}, it could be interesting to find a direct path in microscopic models to not only realize the FCI-SF transition, but to have the translational symmetry further broken from the SF state, leading to the SS state.
The third, slightly more complicated scenario would be a FCI-CDW-SS sequence, where the roton instability triggers the translational symmetry breaking to form a CDW state out of FCI and then the boson condensation further breaks the U(1) symmetry of bosons in CDW, resulting in the SS state with coexisting SF and CDW orders.

In this work, we discover an exotic realization of the third scenario,  by studying the flat band model on the checkerboard lattice with hard-core bosons, where the $\nu=1/2$ FCI states are found to be robust~\cite{DNSheng2011boson, Luo2020boson}. 
Through large-scale density matrix renormalization group (DMRG)~\cite{White1992_DMRG} and exponential tensor renormalization group (XTRG) simulations~\cite{Chen2018XTRG}, by tuning repulsive interactions, we find an intermediate CDW state between FCI and SS, instead of a  direct FCI-SS transition reported in Ref.~\cite{DNSheng2011boson} by exact diagonalization (ED) and Ref.~\cite{Luo2020boson} by infinite DMRG. 
More importantly, we find this emerging CDW state is {\it gapless} but not SF, which we dub as the gapless boson density wave (GBDW). Further increasing the interaction strength, the system transits into an SS state.
While the possible SS state here has been reported, its defining features have not been correctly revealed in previous works~\cite{DNSheng2011boson, Luo2020boson}. 
First, for the translational symmetry breaking, we find the CDW wave vector of the SS state is at ($\pi,\pi$) [denoted as $\rho_{(\pi,\pi)}$] for both sublattices, instead of existing in only one sublattice with the particles on the other sublattice uniformly distributed~\cite{Luo2020boson}, and we will explain that the incorrect CDW order of the previous work may be due to the limited bond dimension in DMRG simulations~\cite{Luo2020boson}. 
Besides, we find the off-diagonal long range order of this SS state resemble the Larkin-Ovchinnikov (LO) type superconductor~\cite{LO1964}, with bosons condensed at $\pm(\frac{\pi}{2},\frac{\pi}{2})$ [denoted as $b_{\pm(\frac{\pi}{2},\frac{\pi}{2})}$], which has not been identified in the previous works~\cite{DNSheng2011boson, Luo2020boson}. Therefore, one shall regard such an SS state as an unconventional {\it finite-momentum} SS state, different from the widely studied ones with simply coexisting orders.

Interestingly, the intertwined CDW order $\rho_{(\pi,\pi)}$ in the SS state can be treated as an induced order from the LO-type SF order $b_{\pm(\frac{\pi}{2},\frac{\pi}{2})}$. When the quantum fluctuations are enhanced from the SS state towards FCI, the  $b_{\pm(\frac{\pi}{2},\frac{\pi}{2})}$ order is partially melted while its product  $\rho_{(\pi,\pi)}\propto b^\ast_{(\frac{\pi}{2},\frac{\pi}{2})}b_{-(\frac{\pi}{2},\frac{\pi}{2})}$ survives, leading to the GBDW state.
This quantum melting resonates profoundly with the physics of thermal melting in pair-density-wave superconductors~\cite{Berg2009pdw, Fradkin2015interwined, Wang2020PDW}, where the intertwined orders are key to understanding the rich phase diagrams and the other induced orders, besides CDW, often include charge-4e superconductors, spin density waves, and nematic orders.
Therefore, the SS to GBDW transition in our work is clearly of the vestigial nature as the symmetry breaking and restoration of the Hamiltonian are in a stepwise manner. 
The vestigial transitions, considering not only competing but also intertwined orders, provide more perspectives on quantum phase transitions and have attracted broad attention~\cite{Nie2017vestigial, Fernandes2019Vestigial, wangVesigial2021,sun2023vestigial}.
What is totally different between our work and the previous vestigial transitions is that, the full quantum melting (at zero temperature) of the Landau symmetry-breaking orders here does not lead to normal states but the FCI state with topological order. 

Meanwhile, the GBDW state itself is intriguing, due to its gapless but non-SF nature, and it is possibly similar to the Bose metal state in this sense. 
The exotic Bose metal has been widely and intensively studied since it was proposed~\cite{Philip2003bosemetal, Philip2019bosemetal}, and it is believed to possibly exist as an intermediate phase between superconductors and Mott insulators instead of a direct transition~\cite{Jaeger1989transitionSI, Fisher1990transition}, with many experimental signatures of dissipative transport properties~\cite{Tsen2016metal, Tamir2019metal, Yang2019BoseMetal}. 
For example, the charge carriers of such Bose metal state in yttrium barium copper oxide (YBCO) are shown to be of two electrons, as Cooper pairs without phase coherence~\cite{Yang2019BoseMetal}.
Theoretically, different systems have been proposed to study the Bose metal, such as the superconducting grain models~\cite{Feigel'man1979spinglass, wagenblast1997transition, das1999metal, Spivak2001transition}, harmonic baths coupled to 1D hard-core bosons~\cite{Cai2014liquid}, moatband models (where the energy minima of the band constitute a degenerate dispersion in
momentum space) in the extreme dilute limit~\cite{sur2019metallic}, geometric frustration based on free bosons~\cite{Hegg2021metal} and lattice-models with ring-exchange interactions~\cite{Paramekanti2002metal}.
Noteworthily, there exist robust numerical simulations of the Bose metal phase in the ring-exchange models on multi-leg ladders~\cite{Sheng2009metal, Block2011metal, Mishmash2011metal, Jiang2013metal}, based on DMRG~\cite{White1992_DMRG} and variational Monte Carlo~\cite{McMillan1965vmc, Ceperley1977vmc} methods, and we also notice the recent numerical discovery of 2D Bose metals in an anisotropic Hubbard model on a square lattice~\cite{cao2024metal, cao2024metal2}, based on the constrained path quantum Monte
Carlo method~\cite{Zhang1997CPMC}. 
While the instability to CDW orders has been discussed~\cite{Paramekanti2002metal, Mishmash2011metal}, to the best of our knowledge, realization of  Bose metals coexisting with CDW orders at zero temperature which are similar to the GBDW state discovered in our work has not been put forward, except the signature of CDW ordering in the variational Monte Carlo wave functions studying the ring-exchange model~\cite{Mishmash2011metal}.
Moreover, the charge order of the GBDW state can not only be seen as a vestige from the SS state, and we will show that the translational symmetry breaking from the FCI side is triggered by the instability of the magneto-roton mode, 
similar to the transition from the fermionic $\nu=2/3$ FCI to the metallic state with the same $\rho_{(\pi,\pi)}$ CDW order (from roton condensation) on the checkerboard lattice~\cite{LHY2024_roton_transition}.
When the CDW order in the GBDW state becomes strong enough, the corresponding SF fluctuation $b_{\pm(\frac{\pi}{2},\frac{\pi}{2})}$ is further triggered and results in the finite-momentum SS state mentioned above, which might suggest incompatibility between fully polarized CDW and bosonic metallic states.

We note, the symmetry-broken bosonic metallic state has been reported in other superconductors that break multiple symmetries, such as the thermal melting of the $a_1+ia_2$ ($a_{1,2}=s,p,d$...) $U(1)\times Z_2$ superconductor~\cite{Babaev2004metal,Bojesen2013metal,Bojesen2014metal,Babaev2015thermal,Yerin2017sis}. Beyond the mean-field level of a direct finite-temperature transition from superconductors to symmetric normal states, there exists intermediate-temperature metallic bosonic state with broken time-reversal $Z_2$ symmetry~\cite{Babaev2004metal,Bojesen2013metal,Bojesen2014metal}, and the experimental signatures have been observed in hole-doped Ba$_{1-x}$K$_x$Fe$_2$As$_2$~\cite{Babaev2021metal}.
This further suggests the connection of the symmetry-broken bosonic metallic state and multi-step transitions, although it is still from the thermal fluctuations instead of quantum fluctuations at $T=0$ in our work.

\begin{figure*}[htp!]
	\centering	
	\includegraphics[width=0.82\textwidth]{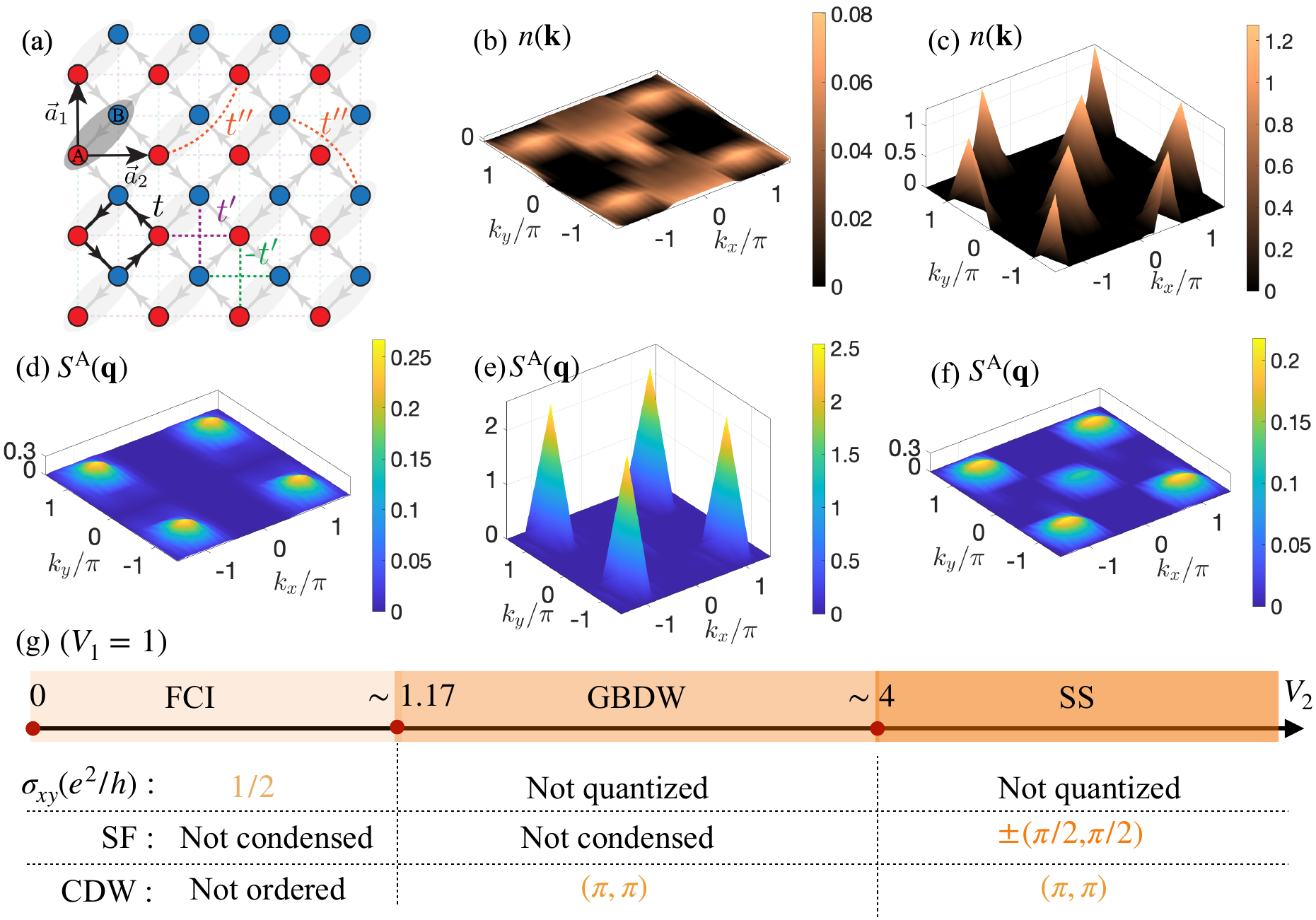}
	\caption{\textbf{Checkerboard lattice model and phase diagram.} (a) Checkerboard lattice with the primitive vectors $\mathbf{a}_1= (0, 1)$ and $\mathbf{a}_2=(1, 0)$. Different hoppings are denoted by different colors and the arrows represent the directions of the loop current. The boson occupation numbers in the momentum space $n(\mathbf{k})$ are shown for (b) GBDW ($V_1=1$, $V_2=1.5$) and (c) SS ($V_1=1$, $V_2=6$) states. The static structure factors of boson density correlation $S^\mathrm{A}(\mathbf{q})$ are shown for (d) FCI ($V_1=1$, $V_2=1$), (e) GBDW ($V_1=1$, $V_2=1.5$) and (f) SS ($V_1=1$, $V_2=6$) states.
		Panels (b-f) are from DMRG results of $4\times16\times2$ cylinders.
		The ground-state phase diagram with fixed $V_1=1$ and tuning $V_2$ is shown in (g). 
		When $V_2\textless1.17$, the ground state is the $\nu=1/2$ FCI with $\sigma_{xy}=1/2$ in units of $e^2/h$ and without other symmetry-breaking orders.
		At intermediate $V_2$, the ground state is a GBDW (gapless) with CDW order $\rho_{(\pi,\pi)}$ but no SF order.
		When $V_2$ is further increased, the ground state is a finite-momentum SS with both LO-type SF order $b_{\pm(\frac{\pi}{2},\frac{\pi}{2})}$ and CDW order $\rho_{(\pi,\pi)}$.
	}
	\label{fig_fig1}
\end{figure*}

Apart from the interesting FCI-GBDW-SS sequence of transitions, a direct FCI-SS transition would still be intriguing. In this work, we also study the similar flat-band Haldane model on honeycomb lattice with hard-core bosons and tuning interactions~\cite{HaldaneModel1988, DNSheng2011boson,Luo2020boson}, where the previous work proposed a direct $\nu=1/2$ FCI-SS transition~\cite{Luo2020boson}. According to our simulations, the SS states do not exist here, instead there is an FCI-Solid I-Solid II sequence of transitions (both solids are incompressible).  
Interestingly, this sequence is triggered by the progressive softening of multiple roton modes in FCI, in that, the minimal roton mode in FCI gives rise to Solid I, and when this CDW is strong enough, a higher-energy mode in FCI goes soft and renders the further doubled translation period of Solid II. Moreover, the Solid I ($\rho_{\frac{\mathbf{b_1}}{2}}$) with stronger quantum fluctuations is also a secondary and {\it vestige} of Solid II ($\rho_{\pm(\frac{\mathbf{b_1}}{4}+\frac{\mathbf{b_2}}{2})}$) with weaker quantum fluctuations. 
Compared to previous studies of softening only a single roton mode~\cite{Mukherjee2022_rotoncondense, Kumar2022_condense,  LHY2024_roton_transition, LHY2024_FQAHS_transition}, this multi-roton softening is absent in the previous literature, let alone the vestigial transitions of the two solids.
\\
\indent We therefore believe our work not only provides more perspectives on the FCI-related transitions, the possible realization of translation-symmetry-broken gapless bosonic states
at zero temperature, the vestigial transitions and intertwined symmetry-breaking orders in bosonic systems, etc, %which paves the way for future studies, 
but also further exhibits the strongly correlated topological flat-band models as appropriate synthetic platforms for exotic quantum states and phase transitions.
\section{FCI-GBDW-SS transition sequence on checkerboard lattice} 
\label{sec_CB}

\subsection{Model and phase diagram}

In this section, we revisit the flat-band model on checkerboard lattice with hard-core bosons half-filled the flat band:
\begin{equation}
	\begin{aligned}
		H =&-\sum_{\langle i,j\rangle}te^{\mathrm{i}\phi_{ij}}(b_i^\dagger b^{\ }_j+\mathrm{H.c.})-\sum_{\langle\hskip-.5mm\langle i,j \rangle\hskip-.5mm\rangle}t'_{ij}(b_i^\dagger b^{\ }_j+\mathrm{H.c.})\\
		&-\sum_{\langle\hskip-.5mm\langle\hskip-.5mm\langle i,j \rangle\hskip-.5mm\rangle\hskip-.5mm\rangle} t''(b_i^\dagger b^{\ }_j+\mathrm{H.c.})
		+V_1\sum_{\langle i,j \rangle}n_in_j
		+V_2\sum_{\langle\hskip-.5mm\langle i,j \rangle\hskip-.5mm\rangle}n_in_j,
	\end{aligned}
	\label{eq:eq1}
\end{equation}
where $b_i^\dagger$($b^{\ }_i$) creates (annihilates) a hard-core boson at the $i$-th site and $n_i=b_i^\dagger b^{\ }_i$,
with nearest-neighbor (NN, $t$), next-nearest-neighbor (NNN, $t'$), and next to next nearest-neighbor (NNNN, $t''$) hoppings, NN repulsive interaction ($V_1$) and NNN repulsive interaction ($V_2$), as shown in Fig.~\ref{fig_fig1} (a). 
The tight-binding parameters are: $t=1$ (as the energy unit), $t'_{ij}=\pm 1/(2+\sqrt{2})$ with alternating sign in edge-sharing plaquettes, $t''=-1/(2+2\sqrt{2})$ and $\phi_{ij}=\frac{\pi}{4}$ along the direction of the arrows.
This paradigmatic flat-band model was studied by ED to hold robust bosonic FCI at $\nu=1/2$ and possibly other fillings with even denominator~\cite{DNSheng2011boson}. 
The $\nu=1/2$ FCI ground state was further confirmed by iDMRG simulations that measured the entanglement spectral
flows and the resultant charge pumping features~\cite{Luo2020boson}.
Apart from the FCI, the previous studies also searched for other competing states within the $V_1-V_2$ phase diagram.
However, the defining features of other states (including SS) still remain elusive (and the GBDW state is completely missed), 
which require more systematic simulations and analyses. As we will show in the present work, 
such efforts turn out to provide new and more interesting physics from this paradigmatic model.

In this work, we focus on $\nu=1/2$ hard-core bosons and fixed $V_1=1$ with tuning $V_2$, and the main results are based on DMRG simulations of $N_y=4$ cylinders, while we also consider larger $N_y=6$ as well as using other numerical methods such as XTRG for thermodynamic properties, with more details of the methods in section~\ref{sec_method}.
The quantum phase diagram is shown in Fig.~\ref{fig_fig1} (g), with the momentum-space boson occupation numbers $n(\mathbf{k})=n_\alpha(\mathbf{k})=\frac{1}{N}\sum_{i,j}e^{-i\mathbf{k}\cdot\mathbf{r}_{i,j}}\langle b^\dagger_{i,\alpha}b^{\ }_{j,\alpha} \rangle$ (where $\alpha=\mathrm{A/B}$ refers to the sublattices, since the results are the same for both sublattices, we mainly present the results on A sublattice), and static structure factors of boson density correlation function $S^\mathrm{\alpha}(\mathbf{q})=\frac{1}{N}\sum_{i,j}e^{-i\mathbf{q}\cdot\mathbf{r}_{i,j}} (\langle n_{i,\alpha}n_{j,\alpha}\rangle-\langle n_{i,\alpha}\rangle \langle n_{j,\alpha}\rangle)$ (we use $\alpha=\mathrm{A}$ sublattice for demonstration if not specified) plotted for different phases.
When $V_2$ is small, the ground state is the $\nu=1/2$ FCI with quantized Hall conductivity but without other broken symmetries.
The structure factor of FCI at $V_1=V_2=1$ is shown in Fig.~\ref{fig_fig1} (d), with a broad peak at ($\pi,\pi$), which refers to the magneto-roton mode~\cite{GMP1985, GMP1986, Yang2012roton, LHY2024thermoFCI,liu2024geometric}.

These observations and the following considerations justify our choice of fixed $V_1=1$ while tuning $V_2$: (1) while the roton-triggered translation symmetry breakings are well studied in fermionic FCI~\cite{LHY2024_roton_transition, LHY2024_FQAHS_transition}, the related knowledge of bosonic FCI is still lacking.
(2) Previous works reported possible SS phase with $\rho_{(\pi,\pi)}$ CDW order at large $V_2$~\cite{DNSheng2011boson,Luo2020boson}, whose wave vector is the same as that of the roton in FCI.
However, according to our simulations, instead of a direct FCI-SS transition in previous works~\cite{DNSheng2011boson,Luo2020boson}, we find an intermediate GBDW state, which is gapless but without SF order.
As shown in Fig.\ref{fig_fig1} (b) and (e), the momentum-space boson occupation number $n(\mathbf{k})$ has no peak, while the structure factor of boson density correlation $S(\mathbf{q})$ has a sharp peak at ($\pi,\pi$), suggesting the CDW order of the GBDW state.
The SS state with $\rho_{(\pi,\pi)}$ order only exists at large $V_2$, and we find the SF order is of LO type, with bosons condensed at $\pm(\pi/2,\pi/2)$ (which to our best knowledge is not reported in any previous works).

\begin{figure}[htp!]
	\centering	
	\includegraphics[width=0.5\textwidth]{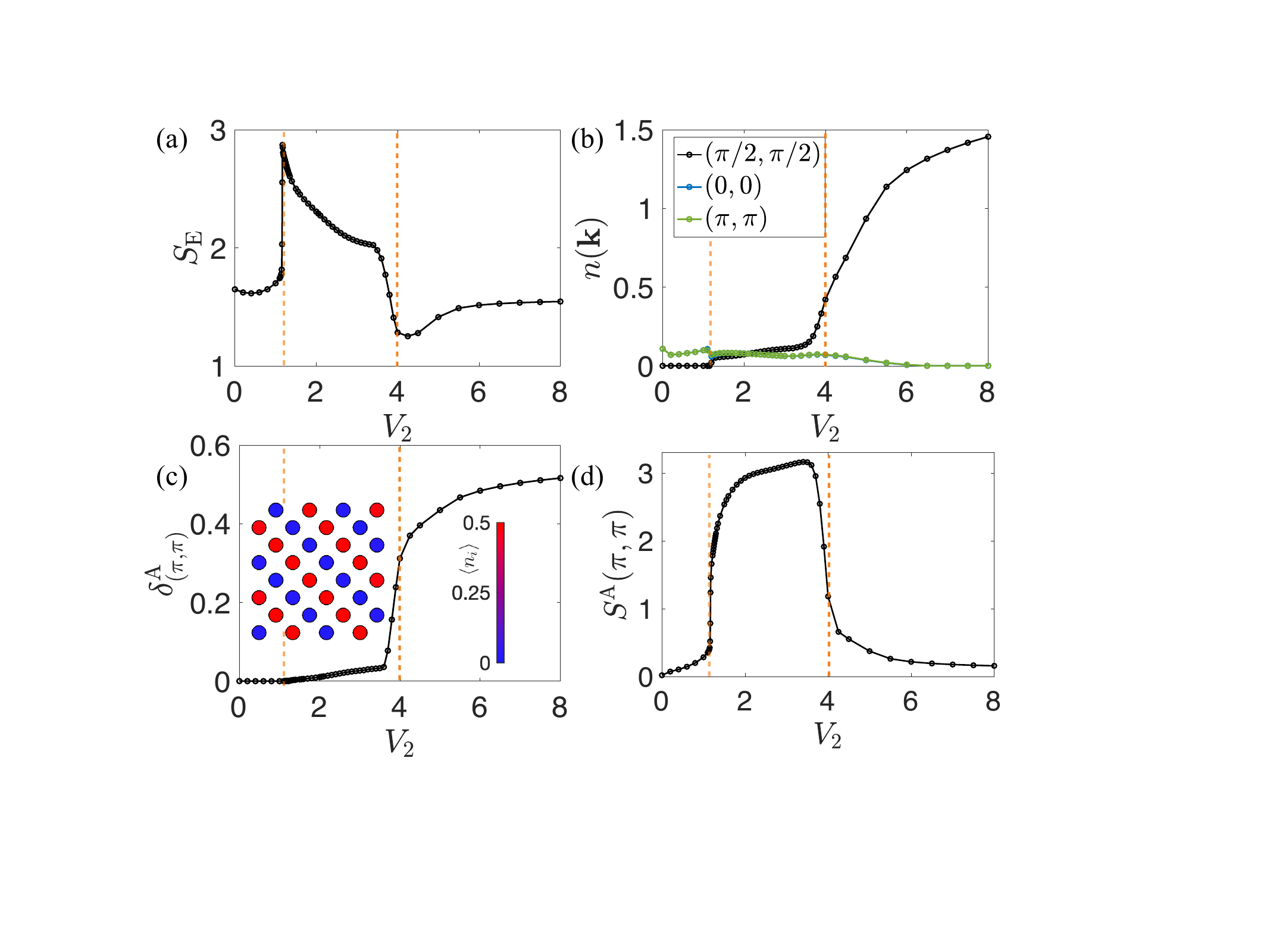}
	\caption{\textbf{The FCI-GBDW-SS sequence of transitions.} DMRG simulation results on $4\times16\times2$ cylinders including, (a) the bipartite entanglement entropy $S_\mathrm{E}$,
		(b) boson occupation number $n(\mathbf{k})$ at different momenta, (c) the real-space order parameter $\delta^\mathrm{A}_{(\pi,\pi)}$ of the CDW order, and (d) the static structure factor $S^\mathrm{A}{(\pi,\pi)}$ of boson density correlation, as functions of $V_2$. 
		The dashed lines represent the ground-state phase boundary as shown in Fig.~\ref{fig_fig1} (g).
		The inset of subfigure (c) is the real-space occupation numbers in the bulk of SS at $V_2=6$, showing the $\rho_{(\pi,\pi)}$ orders are the same for both sublattices.
	}
	\label{fig_fig2}
\end{figure}

To further demonstrate the quantum phase diagram, we show (a) the bipartite entanglement entropy $S_E$, (b) boson occupation number $n(\mathbf{k})$ at different momenta, (c) the real-space order parameter $\delta_{(\pi,\pi)}$ of the CDW order, and (d) the static structure factor $S(\mathbf{q})$ at ($\pi,\pi$), as functions of $V_2$ in Fig.~\ref{fig_fig2}.
The behavior of entanglement entropy supports the sequence of FCI-GBDW-SS transitions in a two-step manner.
The evolution of momentum space boson occupation numbers in Fig.~\ref{fig_fig2} (b) further distinguishes the GBDW from the SS. 
We choose $(0,0)$, $(\pi/2,\pi/2)$, and $(\pi,\pi$) as examples. In GBDW state, the momentum-space density distributions of these points are almost equal, which is also shown in Fig.~\ref{fig_fig1} (b).
When approching the transition point into the SS state, the occupation number at $(\pi/2,\pi/2)$ gradually increases while those at other momenta such as ($0,0$) and ($\pi,\pi$) decreases, and finally the bosons condense at $(\pi/2,\pi/2)$ and the occupation numbers at other points converge to $0$. 
The GBDW-SS transition is possibly as continuous as the evolutions of $n(\mathbf{k})$.

To clarify the CDW order, apart from $S^\mathrm{A}(\pi,\pi)$, we define the real-space order parameter as $\delta^\mathrm{\alpha}_{(\pi,\pi)}=\frac{2}{N_xN_y}\sum_{i,j}(-1)^{i+j}\langle n^\alpha_{i\mathbf{a_1}+j\mathbf{a_2}}\rangle$, shown in Fig.~\ref{fig_fig2} (c). 
In the SS state, the translational symmetry is broken in real space, so that the $(\pi,\pi)$ fluctuations decrease, as shown in Fig.~\ref{fig_fig2} (d). In the GBDW state, the real-space boson occupations seem uniform, while the structure factors show high peaks, this is because the wavefunction obtained by DMRG is a `cat' state (superposition of different degenerate CDW patterns) while the density-density correlations still manifest the long-range order, which we will explain more in Sec.~\ref{sec:secIIB}.
Besides, as shown in the inset of Fig.~\ref{fig_fig2} (c), the SS state has the same translational symmetry breaking in both sublattices, the same for GBDW, in contrast to the previous iDMRG result~\cite{Luo2020boson} which found the $\rho_{(\pi,\pi)}$ order of SS exists in only one sublattice (previous ED work did not focus on this~\cite{DNSheng2011boson}).
We find the incorrect CDW order (in only one sublattice) is due to the limited bond dimensions in iDMRG simulations, which we will show more detailed comparison in the Appendix~\ref{appendix_a}.

This sequence of FCI-GBDW-SS transition is rather intriguing.
From the SS side, one sees when $V_2$ decreases, the increasing quantum fluctuation melts only the SF order $b_{\pm(\frac{\pi}{2},\frac{\pi}{2})}$ , while its product---the CDW order---survives, resulting in the vestigial GBDW phase without SF. 
This SS-GBDW transition is similar to the vestigial transitions in pair-density-wave superconductors ~\cite{Berg2009pdw, Fradkin2015interwined, Wang2020PDW}.
While the vestigial transitions from thermal fluctuations are better studied, such a SS-GBDW scenario at $T=0$ in our work paves the way for further understanding the vestigial transitions and intertwined orders in SS.
From the FCI side of the phase diagram, the translation-symmetry breaking of the GBDW state is triggered by the softening of roton mode~\cite{Mukherjee2022_rotoncondense, Kumar2022_condense,LHY2024_roton_transition,LHY2024_FQAHS_transition}.
As shown in Fig.~\ref{fig_fig2} (d), when approaching the transition point from FCI, the structure factor at $(\pi,\pi)$ continues to increase, referring to the closing of roton gap. We will discuss the corresponding thermodynamic signature of this process in Sec.~\ref{sec:secIID}.
After the transition into the GBDW, $S^\mathrm{A}(\pi,\pi)$ becomes very high and the long-range CDW order establishes, although the density fluctuations at ($\pi,\pi$) do not go down since the real-space boson distribution is still uniform due to the superposition of different CDW patterns.
Besides, when the CDW order is strong enough, it results in the LO-type SF order, which is different from other roton-condensation scenarios (at either zero or finite-momentum) in the literature of FCI/FQH states~\cite{GMP1986,  Regnault2017nematic, Mukherjee2022_rotoncondense, Kumar2022_condense, puMicroscopic2024,  LHY2024_roton_transition, LHY2024_FQAHS_transition}. 
Considering the gapless, translational-symmetry-breaking, and non-SF nature of the intermediate GBDW state, this scenario of the two-step transition becomes more exotic.

\subsection{The GBDW state}
\label{sec:secIIB}
\begin{figure}[htp!]
	\centering	
	\includegraphics[width=0.5\textwidth]{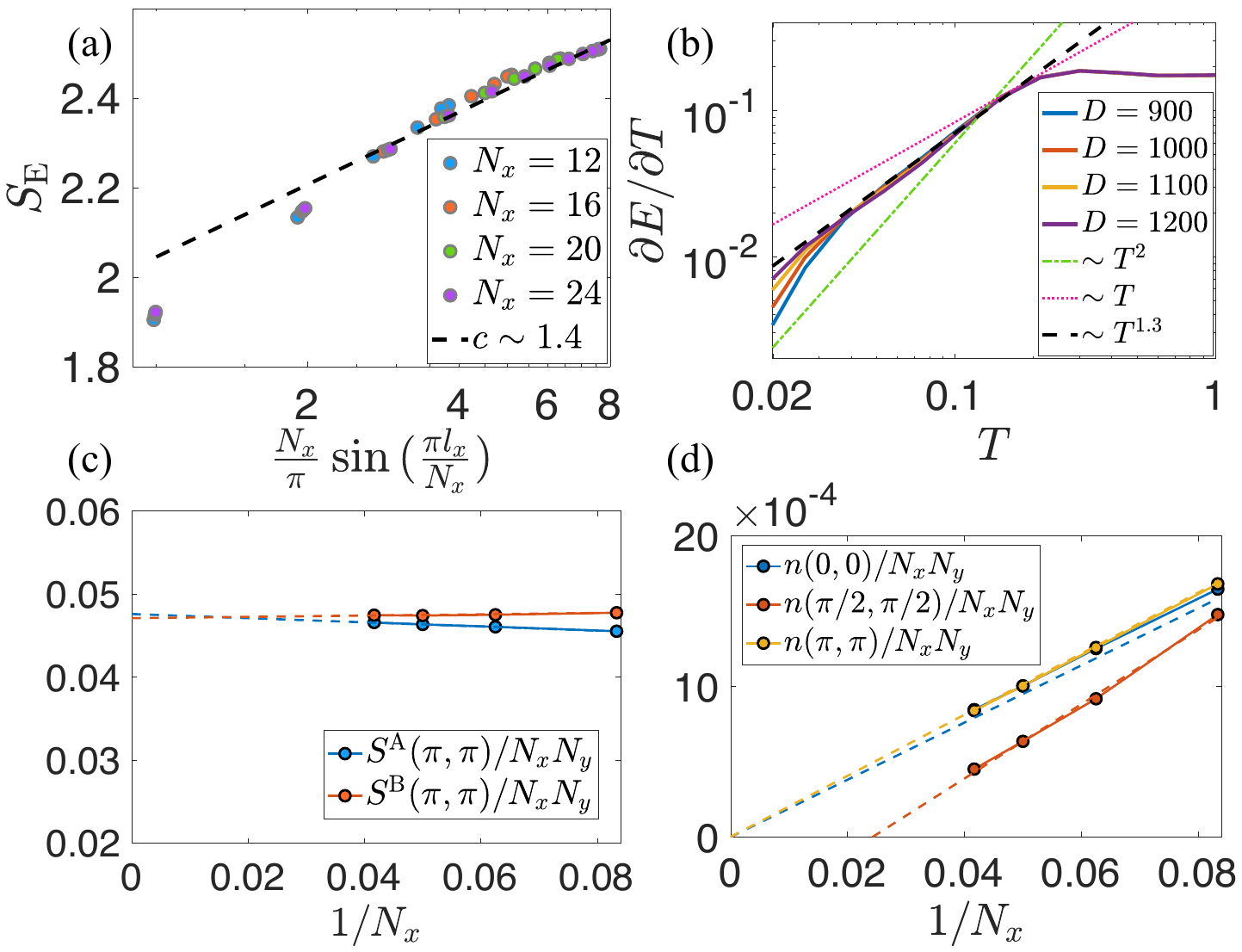}
	\caption{\textbf{The GBDW state from $N_y=4$ cylinders at $V_1=1$ and $V_2=1.6$.} (a) Entanglement entropy versus the conformal distance (defined as $\frac{N_x}{\pi}\sin{(\frac{\pi l_x}{N_x})}$) from cylinders up to $N_x=24$.
		The dashed line is fitted from entanglement entropy using the formula $S_\mathrm{E}=\frac{c}{6}\mathrm{ln}(\frac{N_x}{\pi}\sin{(\frac{\pi l_x}{N_x})})+g$, where $g$ is fitting constant and the fitted central charge is $c\sim1.4$.
		(b) The specific heat from XTRG simulations of a $4\times12\times2$ cylinder with bond dimension up to $D=1200$. At low temperature, the scaling behavior of specific heat is converging to a power-law decay with a power around $1.3$ denoted by the black dashed line.
		We also plot linear ($\sim T$) and square ($\sim T^2$) decay as dotted lines for comparison.
		(c) Finite-size scaling of structure factors of boson density correlation of both sublattices ($S^\mathrm{A/B}(\pi,\pi)$).
		(d) Finite-size scaling of momentum-space boson occupation number at different momenta. The dashed lines in panels (c,d) are fitted from the rescaled data (divided by the number of unit cells $N_x N_y$) as functions of $1/N_x$.
	}
	\label{fig_fig3}
\end{figure}

As introduced above, the GBDW state is gapless, with broken translational symmetry but no SF order. To further demonstrate these non-trivial features, more detailed results are shown in Fig.~\ref{fig_fig3} (we still focus on $N_y=4$ cylinders and take $V_1=1$ and $V_2=1.6$ as an example). 
We simulate the bipartite entanglement entropy 
to confirm the gapless nature of the GBDW state in the quasi-one-dimensional system, and fit the well-converged entanglement entropy up to $N_x=24$ according to the logarithmic correction to the area law as the function $S_\mathrm{E}=\frac{c}{6}\mathrm{ln}(\frac{N_x}{\pi}\sin{(\frac{\pi l_x}{N_x})})+g$, where $g$ is fitting constant.
As shown in Fig.~\ref{fig_fig3} (a), the fitted central charge is $c\sim1.4$. 
Although it is close to 1, the exact value of the central charge still needs to be checked from the scaling of the width of the cylinder.
However, the fitted non-zero central charge here already suggests the gapless feature of this GBDW state.
Supplementary data of entanglement entropy is shown in the Appendix~\ref{appendix_a}. 

We further show the specific heat of GBDW from XTRG simulations of a $4\times12\times2$ cylinder in Fig.~\ref{fig_fig3} (b). With the increase of bond dimension, the low-temperature specific heat is converging to a power-law scaling, and the fitted power is around $\frac{\partial E}{\partial T}\sim T^{1.3}$.
Although the exact scaling might need more accurate thermodynamic simulations and considering maybe even lower temperature, the current power-law decay of the low-$T$ specific heat supports our conclusion that the GBDW state has gapless excitations.

Meanwhile, as shown in Fig.~\ref{fig_fig1} (e) and Fig.~\ref{fig_fig2} (c,d), the real-space boson density is almost uniform in GBDW while the presence of the $\rho_{(\pi,\pi)}$ charge order is manifested from structure factors of boson density correlation $S^\mathrm{A/B}(\pi,\pi)$.
To confirm our result, we show the finite-size scaling of both $S^\mathrm{A/B}(\pi,\pi)/N_xN_y$ in Fig.~\ref{fig_fig3} (c), which suggests the established, finite, and the same value of long-range CDW order parameters of both sublattices in the thermodynamic limit, in agreement with our conclusions.
This is also against the previous work that $\rho_{(\pi,\pi)}$ exists in only one sublattice ~\cite{Luo2020boson} (we will show more detailed comparison that this is due to limited bond dimension in  Appendix~\ref{appendix_a}), as we have shown for the SS state.
Furthermore, the major difference of the GBDW from SS is that there is no SF component, as shown in Fig.~\ref{fig_fig1} (b) and  Fig.~\ref{fig_fig2} (b) that the momentum-space boson occupation is almost uniform in different momenta.
We show the finite-size scaling of $n(\mathbf{k})/N_xN_y$ at ($0,0$), ($\pi/2,\pi/2$), and ($\pi,\pi$) respectively in Fig.~\ref{fig_fig3} (d).
It turns out that these values scale almost linearly with $1/N_x$ and extrapolate to $0$ in the thermodynamic limit, supporting that there is no condensation of bosons at these momenta.
Besides, $n(\pi/2,\pi/2)/N_xN_y$ decays even faster than those of other momenta, which further rules out the SF component in the GBDW phase, as in SS bosons condense at $\pm(\pi/2,\pi/2)$.

\begin{figure}[htp!]
	\centering	
	\includegraphics[width=0.5\textwidth]{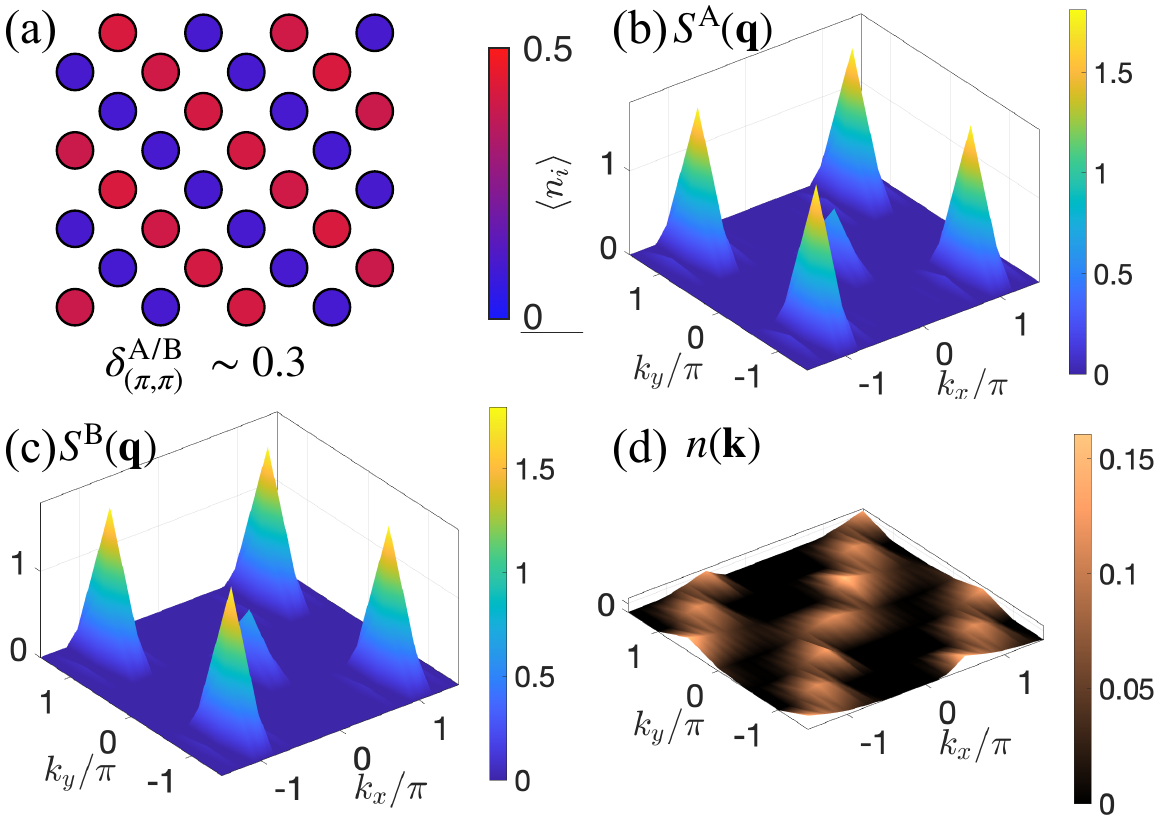}
	\caption{\textbf{The GBDW state under a small pinning field $h=0.001$ from $4\times16\times2$ cylinders at $V_1=1$ and $V_2=1.5$.} (a) The real-space boson distribution in the bulk of the cylinder with the order parameters of both sublattices $\delta^\mathrm{A/B}_{(\pi,\pi)}\sim0.3$. The structure factors of boson density correlations of both sublattices ($S^\mathrm{A/B}(\mathbf{q})$) are shown in (b) and (c), respectively.  The CDW pattern is the same for both sublattices. (d) The momentum-space boson occupation number $n(\mathbf{k})$, no condensation is observed. 
	}
	\label{fig_fig4}
\end{figure}

So far the results are obtained from the Hamiltonian in Eq.~\eqref{eq:eq1} without any pinning field, and we find the almost uniform real-space boson density distribution is due to the superposition of degenerate CDW patterns.
To further confirm this conclusion, we consider adding a very small pinning field of $\rho_{(\pi,\pi)}$ order, defined as $H_{\rho_{(\pi,\pi)}}=-h\sum_{i,j}(-1)^{\alpha'}(-1)^{i+j}(n^\alpha_{i\mathbf{a_1}+j\mathbf{a_2}})$, where $\alpha'=\pm1$ for $\mathrm{A/B}$ sublattices respectively (this is to pin one of the degenerate patterns as in the inset of Fig.~\ref{fig_fig2} (c) for SS).
The results of adding a very small pinning field with $h=0.001$ on a $4\times16\times2$ cylinder at $V_1=1$ and $V_2=1.5$ are shown in Fig.~\ref{fig_fig4}.
In this case we find the translation symmetry breaking is explicitly seen from the real-space boson distribution in Fig.~\ref{fig_fig4} (a), with the order parameters of both sublattices $\delta^\mathrm{A/B}_{(\pi,\pi)}\sim0.3$.
In addition, we check and compare the bipartite entanglement entropies in the cases of $h=0$ and $h=10^{-3}$, with the difference around $\log2$,
further supporting our conclusion that the results of GBDW without pinning field are the superpositions of degenerate patterns. 
Besides, the pinned order parameter of $\rho_{(\pi,\pi)}$ is smaller than those of SS, and the order in GBDW will continue to increase towards the GBDW-SS transition point [the increasing $S^\mathrm{A}(\pi,\pi)$ in Fig.\ref{fig_fig2} (d)], which suggests that, the SF $b_{\pm(\frac{\pi}{2},\frac{\pi}{2})}$ order is triggered from the $\rho_{(\pi,\pi)}$ order when it becomes strong enough in large $V_2$ region.
On the other hand, from large to small $V_2$, the increasing quantum fluctuations first melt the LO-type $b_{\pm(\frac{\pi}{2},\frac{\pi}{2})}$ order, while its product  $\rho_{(\pi,\pi)}\propto b^\ast_{(\frac{\pi}{2},\frac{\pi}{2})}b_{-(\frac{\pi}{2},\frac{\pi}{2})}$ survives and becomes weaker gradually.
Moreover, with the small pinning field and the explicit real-space order parameters, the structure factors $S^\mathrm{A/B}(\pi,\pi)\sim1.8$ both become weaker as compared to Fig.~\ref{fig_fig1} (e) ($S^\mathrm{A}(\pi,\pi)\sim2.5$), as shown in Fig.~\ref{fig_fig4} (b,c) respectively, since the CDW fluctuations would naturally become weaker when the corresponding order parameters establish.
In the structure factors of both GBDW (whether with pinning field or not) and SS states, we find some signals at ($0,0$), which is possibly due to the broken rotational symmetry.

In Fig.~\ref{fig_fig4} (d), with the same small pinning field and explicit translational symmetry breaking in real-space boson occupations, we find $n(\mathbf{k})$ exist in the anti-diagonal of the Brillouin zone (BZ), slightly different from that without pinning field in Fig.~\ref{fig_fig1} (b). 
However, it is still consistent with our conclusion that there is no SF order in the GBDW state since the momentum-space boson occupation numbers are almost equal along the anti-diagonal of BZ and we observe no peaks of boson condensation.
Furthermore, it also supports our conclusion that the results of GBDW without pinning field are superpositions of degenerate patterns, since the $n(\mathbf{k})$ in Fig.~\ref{fig_fig1} (b) is actually a superposition of $n(\mathbf{k})$ along two diagonals, and 
the $n(\mathbf{k})$ along one diagonal in Fig.~\ref{fig_fig4} (d) is almost twice of that of Fig.~\ref{fig_fig1} (b) since one of the two degenerate diagonals is selected by the small pinning field and $n(\mathbf{k})$ re-weighted.
We will now show further supporting data by considering wider cylinders.

\begin{figure}[htp!]
	\centering	
	\includegraphics[width=0.5\textwidth]{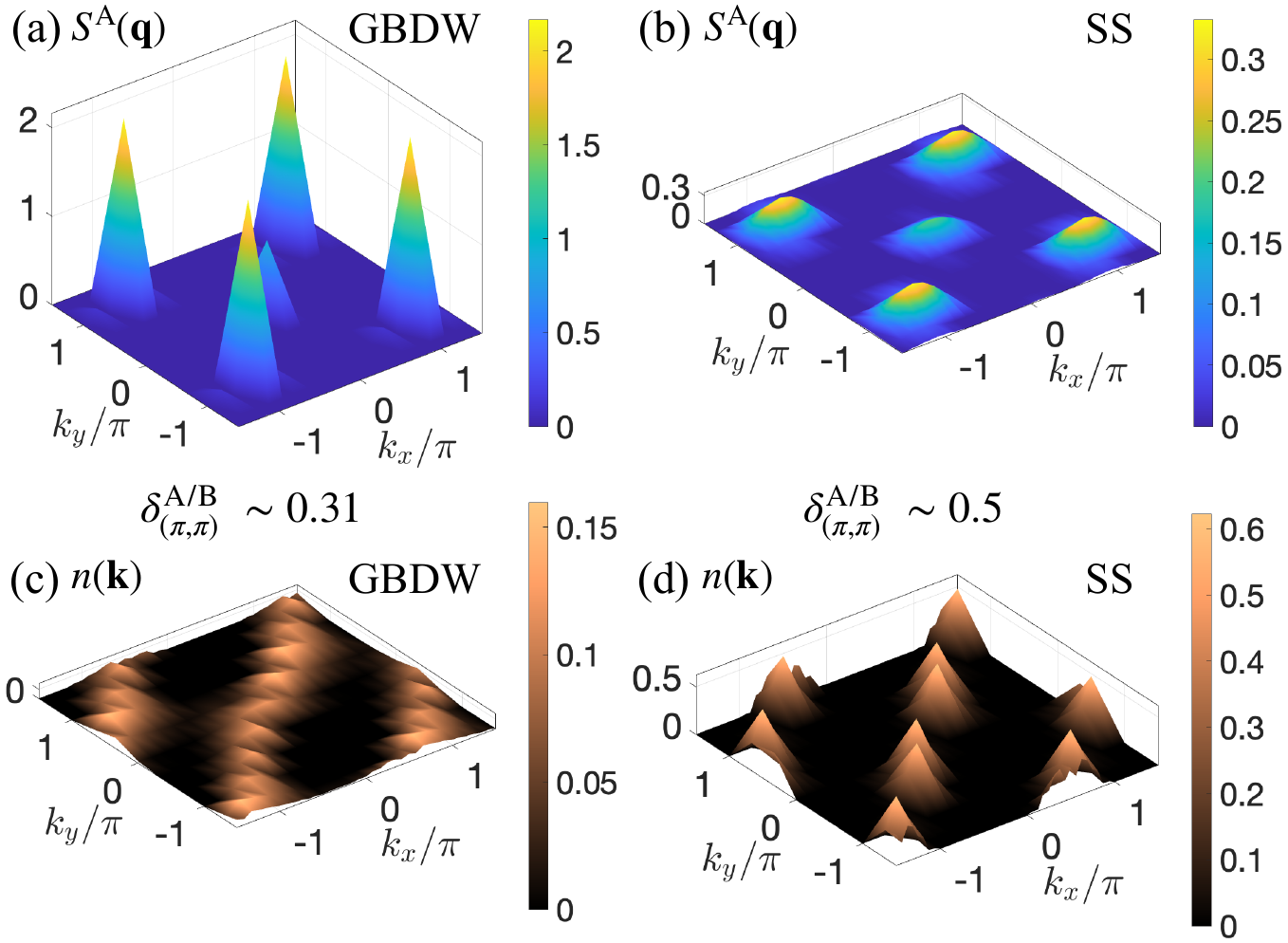}
	\caption{\textbf{The GBDW and SS states from $6\times16\times2$ cylinders.} 
		The structure factors of boson density correlations $S^\mathrm{A}(\mathbf{q})$ are shown for (a) GBDW and (b) SS, respectively.
		The momentum-space boson occupation numbers $n(\mathbf{k})$ are shown for (c) GBDW and (d) SS, respectively.
		We consider $V_1=1$ and $V_2=1.5$ for GBDW, and $V_1=1$ and $V_2=6$ for SS.
		No pinning field is applied here, and the translational-symmetry breaking is seen explicitly from real-space boson distribution as the CDW order parameters $\delta^{A/B}_{(\pi,\pi)}$ of both sublattices are around $0.31$ for GBDW and $0.5$ for SS.
		In the $6\times16\times2$ cylinder without twisting the boundary, $\pm(\pi/2,\pi/2)$ is not considered in the BZ, and the closest momenta to $(\pi/2,\pi/2)$ are $(3\pi/8,\pi/3)$ and $(5\pi/8,2\pi/3)$. 
	}
	\label{fig_fig5}
\end{figure}

\subsection{Results on $N_y=6$ cylinders}

\begin{figure*}[htp!]
	\centering	
	\includegraphics[width=0.84\textwidth]{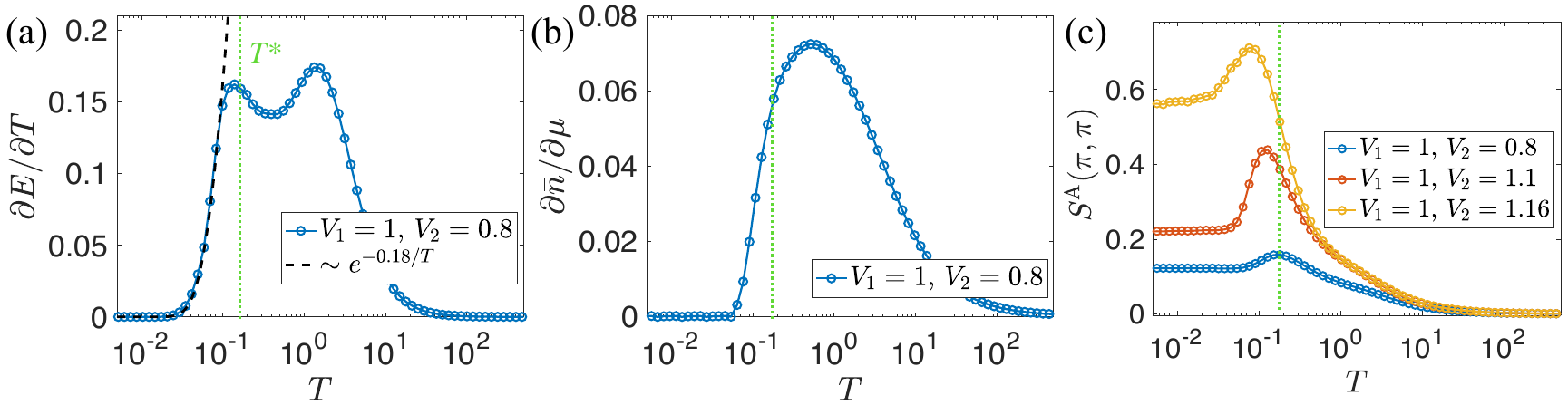}
	\caption{\textbf{The XTRG results of $\nu=1/2$ FCI on $4\times12\times2$ cylinders.} 
		The specific heat $\partial E/\partial T$, compressibility $\partial \bar{n}/\partial\mu$, and structure factor of boson density correlations $S^\mathrm{A}(\pi,\pi)$ at $V_1=1$ and $V_2=0.8$ are shown in (a-c) respectively, and the green dashed line represents the onset temperature of this FCI.
		In panel (a), the specific heat below $T^\ast$ is well fitted to an activation behavior and the activation gap is around $T^\ast\sim0.18$.
		In panel (c), the finite-temperature behaviors of $S^\mathrm{A}(\pi,\pi)$ at some other parameters are shown as well. The roton mode goes soft (the peak in $S^{A}(\pi,\pi)$ moves to lower temperatures and is enhanced) when approaching the FCI-GBDW transition point.
	}
	\label{fig_fig6}
\end{figure*}

In previous sections, our results are based on $N_y=4$ cylinders, and we show results of $N_y=6$ cylinders without any pinning field for GBDW and SS states in Fig.~\ref{fig_fig5}. Although $N_y=6$ is wider, it does not contain $(\pi,\pi)$ and $(\pi/2,\pi/2)$ points simultaneously in the BZ, even if considering any twist boundary conditions.
Therefore, we do not use twisted boundary conditions in our DMRG simulations and the obtained results are still consistent with our conclusions.
In the $N_y=6$ results, the translation symmetry is spontaneously broken and this can be direcly seen from the real-space boson distribution, suggesting the DMRG results have selected one of the degenerate patterns.
In  the GBDW state ($V_1=1$ and $V_2=1.5$) the real-space order parameters of both sublattices are $\delta^\mathrm{A/B}_{\pi,\pi}\sim0.31$, which are almost equal to the result for $N_y=4$ with the same interactions and a small pinning field $h=0.001$ [Fig.~\ref{fig_fig4} (a)].
This further supports that the $N_y=4$ result without pinning field is the superposition of degenerate patterns and a very small pinning field is enough to obtain the exact value of order parameter for one of the degenerate states.
For the SS state ($V_1=1$ and $V_2=6$), the CDW order parameters are $\delta^\mathrm{A/B}_{\pi,\pi}\sim0.5$, which is also consistent with the $N_y=4$ result in Fig.~\ref{fig_fig2} (c). 
Since the CDW order is much stronger in SS, the CDW fluctuation is weaker, so the structure factor $S^\mathrm{A}(\pi,\pi)$ of boson density correlations for SS in Fig.~\ref{fig_fig5} (b) is smaller than that for GBDW in Fig.~\ref{fig_fig5} (a).

The results of momentum-space boson occupations in $N_y=6$ are also in agreement with those in $N_y=4$. 
As shown in Fig.~\ref{fig_fig5} (c), for the GBDW state, the bosons still exist in the anti-diagonal of the BZ with almost equal distribution.
More importantly, apart from the same shape of $n(\mathbf{k})$, the exact value at each momentum point along this anti-diagonal for the $6\times16\times2$ cylinder [Fig.~\ref{fig_fig5} (c)] is approximately the same as that for the $4\times16\times2$ [Fig.~\ref{fig_fig4} (d)] with the same interaction parameters and a very small pinning field for selecting one of the degenerate patterns.
This further suggests that there is no SF in the GBDW state.
Although $\pm(\pi/2,\pi/2)$ are not contained in the $N_y=6$ simulations, the closest two points to $(\pi/2,\pi/2)$ are $(3\pi/8,\pi/3)$ and $(5\pi/8,2\pi/3)$. 
As shown in Fig.\ref{fig_fig5} (d), the high values at $(3\pi/8,\pi/3)$ and $(5\pi/8,2\pi/3)$ still support our conclusion that the bosons in the SS state would condense at $\pm(\pi/2,\pi/2)$, while the values at other momenta tend to converge to $0$.

\subsection{Thermodynamics of FCI and softening of roton mode}
\label{sec:secIID}
The thermodynamics of fermionic FCI and FCI+CDW have been well studied~\cite{LHY2024thermoFCI, LHY2024_FQAHS_PSM}, while such analysis in bosonic systems is lacking. 
Therefore, we fill in this gap here, and the results turn out to be consistent with the fermionic cases.
The XTRG simulations of thermal results of this $\nu=1/2$ FCI on $4\times12\times2$ cylinders are shown in Fig.~\ref{fig_fig6}.
Since the simulations would be simpler and more accurate when getting away from the transition point ($V_2\sim1.17$) between FCI and GBDW, we mainly take $V_1=1$ and $V_2=0.8$ as an example.
From the specific heat [Fig.\ref{fig_fig6} (a)], we observe a clear peak at low temperature $T^\ast\sim0.18$. 
Below $T^\ast$, the low-$T$ specific heat decays exponentially with an activation gap $\sim T^\ast$, which suggests that $T^\ast$ refers to the scale of the lowest gap of this FCI.
In previous parts, we have determined the roton gap of this FCI in our simulations is at $(\pi,\pi)$, as shown from the structure factors [Fig.\ref{fig_fig1} (d) and Fig.\ref{fig_fig2} (d)].
Therefore, we also analyze the structure factor of boson density correlations at finite temperature, as shown in Fig.~\ref{fig_fig6} (c). 
At $T^\ast$ (for $V_1=1$ and $V_2=0.8$), the $S^\mathrm{A}(\pi,\pi)$ shows a peak and gradually converges to a constant at lower temperature where the specific heat shows the activation behavior.
This result further illustrates that $T^\ast$ refers to the lowest gap which is the magneto-roton gap of the FCI.
Besides, this roton gap of charge-neutral excitations is also the onset temperature of incompressible FCI (compressibility fastly drops towards 0 below $T^\ast$), as shown in [Fig.~\ref{fig_fig6} (b)], which is much lower than the estimated charge gap $\Delta_\mathrm{cg}=E(N=96, N_b=25)+E(N=96, N_b=23)-2E(N=96, N_b=24)\sim0.84$ at $T=0$ from DMRG simulations of the same cylinder.
This is consistent with previous thermodynamic studies of FCI that the proliferation of neutral excitations leads to charged excitations and weakens the FCI at temperatures much lower than the charge gap~\cite{LHY2024thermoFCI, LHY2024_FQAHS_PSM}.

\begin{figure*}[htp!]
	\centering	
	\includegraphics[width=0.83\textwidth]{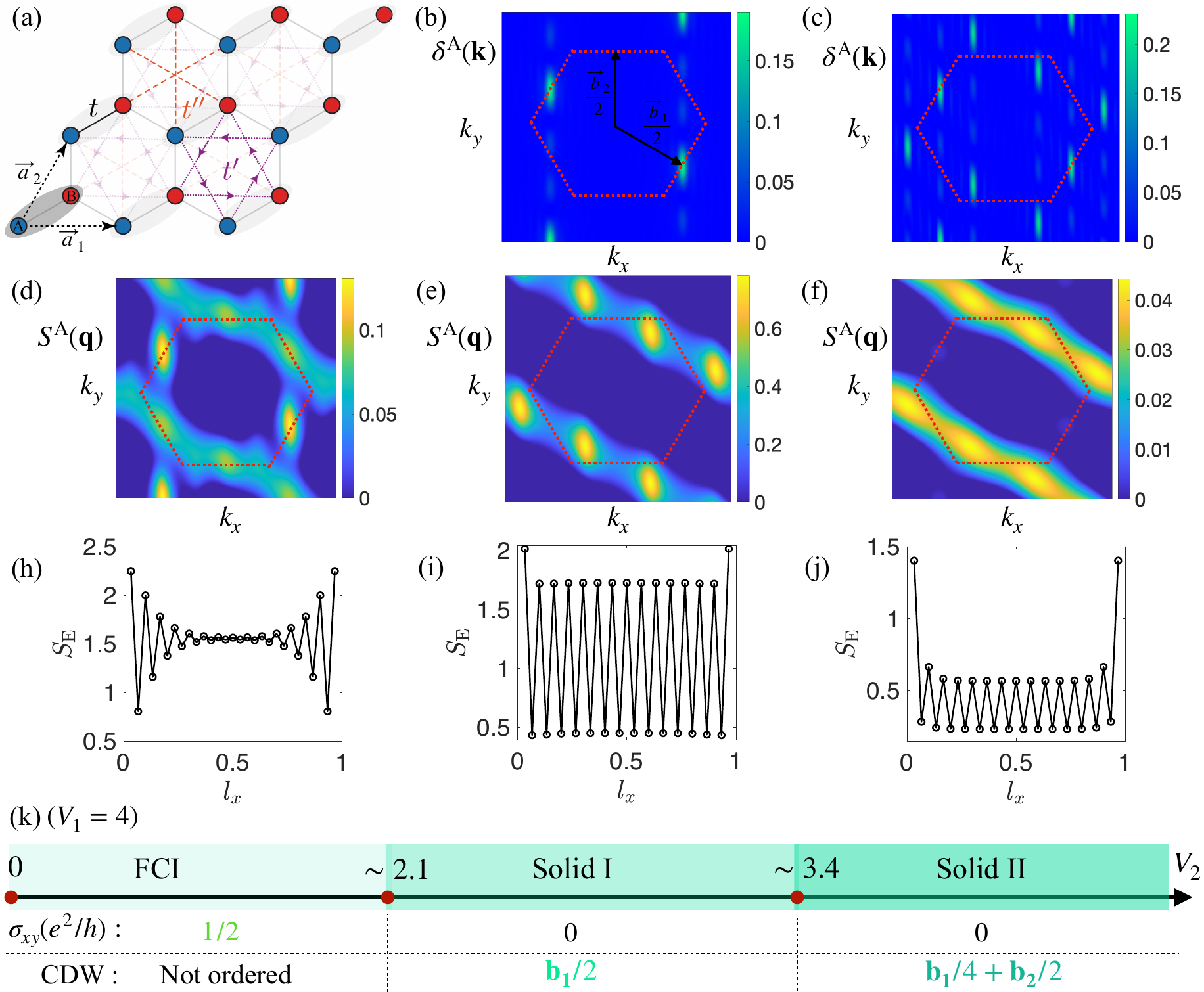}
	\caption{\textbf{Honeycomb lattice model and phase diagram.} (a) Honeycomb lattice with the primitive vectors (black dashed arrows) $\mathbf{a}_1= (1, 0)$ and $\mathbf{a}_2=(1/2, \sqrt{3}/2)$. Different hoppings are denoted by different colors and the purple arrows represent the directions of the loop current of the Haldane model. 
		The Fourier transformations of real-space boson occupation numbers $\delta^\mathrm{A}(\mathbf{k})$ are shown for (b) Solid I ($V_1=4$, $V_2=2.6$) and (c) Solid II ($V_1=4$, $V_2=4$) phases. 
		Half of the reciprocal lattice vectors are labeled by black arrows in panel (b).
		The structure factors of boson density correlation $S^\mathrm{A}(\mathbf{q})$ are shown for (d) FCI ($V_1=4$, $V_2=1.5$), (e) Solid I ($V_1=4$, $V_2=2.6$), and (f) Solid II ($V_1=4$, $V_2=2.6$) states.
		The well-converged entanglement entropy $S_\mathrm{E}$ are shown for the three gapped phases: (h) FCI ($V_1=4$, $V_2=1.5$), (i) Solid I ($V_1=4$, $V_2=2.6$), and (j) Solid II ($V_1=4$, $V_2=2.6$).
		Panels (b-j) are from DMRG results of $N_y=4$ cylinders.
		The ground-state phase diagram with fixed $V_1=4$ and tuning $V_2$ is shown in (k). 
		When $V_2\textless2.1$, the ground state is the $\nu=1/2$ FCI with $\sigma_{xy}=1/2$ in units of $e^2/h$ and without other symmetry-breaking orders.
		At intermediate $V_2$, the ground state is a Solid I state with CDW order $\rho_{\mathbf{b_1}/2}$.
		When $V_2$ is further increased, the translation period is doubled, which leads to a Solid II state  with CDW order $\rho_{\pm(\mathbf{b_1}/4+\mathbf{b_2}/2)}$. 
	}
	\label{fig_fig7}
\end{figure*}

Moreover, we also show the finite-temperature structure factors with larger $V_2$ in Fig.~\ref{fig_fig6} (c). It can be clearly observed that, when approaching the transition point between FCI and GBDW, the peak and low-$T$ converged values of $S^\mathrm{A}(\pi,\pi)$ are getting higher, while the temperature scale of the peak of $S^\mathrm{A}(\pi,\pi)$ is getting lower.
This is consistent with our conclusions that the roton gap at ($\pi,\pi$) is softening and finally brings about the translational symmetry breaking with the same wave vector, although it is hard to identify whether this roton-triggered FCI-GBDW transition is continuous or not from current results.

Before we turn to discuss the results on honeycomb lattice, we would like to mention that another possible SS state with a different CDW order from the one studied in our work, was proposed in the large $V_1$ region of this $\nu=1/2$ phase diagram, according to the previous studies~\cite{DNSheng2011boson, Luo2020boson}.
However, except for the analysis of CDW, whether the state is gapless is unclear. If it is indeed a SS state, the information regarding the type of SF order (such as the momentum point of boson condensation) is not available in previous studies.
Therefore, it might be interesting for future works to revisit that region with more careful analysis, and hopefully exotic physics would appear as we show here, while we just focus on the $\rho_{(\pi,\pi)}$-related scenario in this work.

% ================================ %

% ================================ %
\section{FCI-Solid I-Solid II transition on honeycomb lattice}\label{sec_HC}
In this section, we turn to the flat-band Haldane model on honeycomb lattice~\cite{HaldaneModel1988} with hard-core bosons half-filled the flat band:
\begin{equation}
	\begin{aligned}
		H =&-\sum_{\langle i,j\rangle}t(b_i^\dagger b^{\,}_j+\text{H.c.})-\sum_{\langle\hskip-.5mm\langle i,j \rangle\hskip-.5mm\rangle}t'(e^{i\phi}b_i^\dagger b^{\ }_j+\text{H.c.})\\
		&-\sum_{\langle\hskip-.5mm\langle\hskip-.5mm\langle i,j \rangle\hskip-.5mm\rangle\hskip-.5mm\rangle} t''(b_i^\dagger b^{\ }_j+\text{H.c.})+V_1\sum_{\langle i,j\rangle}n_in_j+V_2\sum_{\langle\hskip-.5mm\langle i,j \rangle\hskip-.5mm\rangle}n_in_j,
	\end{aligned}
	\label{eq:eq2}
\end{equation}
where $b_i^\dagger$ ($b_i$) creates (annihilates) a hard-core boson at the $i$-th site. We consider a zigzag geometry as shown in Fig.~\ref{fig_fig7} (a) and set nearest-neighbor (NN) $t=1$, next-nearest-neighbor (NNN) $t'=0.6$, next-next-nearest-neighbor (NNNN) $t''=-0.58$ and $\phi=0.4\pi$, which are found as optimal flat-band parameters in this model~\cite{DNSheng2011boson}.  $V_1$ ($V_2$) refers to the amplitude of NN (NNN) repulsive interactions. %The lattice model is shown in Fig.\ref{fig_fig7} (a).

In previous works, different phase diagrams were given in this model when focusing on the same parameter path (fixed $V_1=4$ and tuning $V_2$), including a sequence of FCI-SF-Solid transitions~\cite{DNSheng2011boson} and a sequence of FCI-SS-Solid transitions~\cite{Luo2020boson}. 
Considering the rich discoveries overlooked in previous works of the case on checkerboard lattice, we are motivated to carefully investigate the possible exotic transitions in this honeycomb-lattice model, and again find important information missed by previous works.

According to our DMRG simulations on $N_y=4$ cylinders, the updated phase diagram along the same path as in the previous work~\cite{DNSheng2011boson, Luo2020boson} is shown in Fig.~\ref{fig_fig7} (k). 
We find that, although the transition points are close to those in the iDMRG work~\cite{Luo2020boson}, there exists no intermediate SS state. 
Instead, the two states at intermediate $V_2$ and large $V_2$ are both gapped solids with different CDW orders, which we label as Solid I and Solid II, respectively, and the gapped natures of all of the three states are exhibited from the bipartite entanglement entropies as shown in Fig.\ref{fig_fig7} (h-j), where there are well-established plateaus of $S_\mathrm{E}$ when cutting in the bulk of the cylinders.
Moreover, we show the Fourier transformations of the real-space boson occupation numbers
$\delta^\mathrm{A/B}(\mathbf{k})=\sum_j e^{-i\mathbf{k}\cdot\mathbf{r}}(\hat{n}_{j,\mathrm{A/B}}-N_\mathrm{b}/N)/(N_xN_y)$ (the results of A/B sublattices are the same and we show the $\delta^\mathrm{A}(\mathbf{k})$ for instance) for Solid I and Solid II in Fig.~\ref{fig_fig7} (b,c), respectively.
It is clear that in Solid I, the wave vector of the CDW order is at $\mathbf{b_1}/2$.
More interestingly, the CDW wave vector of Solid II is at $\pm(\mathbf{b_1}/4+\mathbf{b_2}/2)$, and the $\rho_{\frac{\mathbf{b_1}}{2}}$ order can be seen as the secondary product of the $\rho_{\pm(\frac{\mathbf{b_1}}{4}+\frac{\mathbf{b_2}}{2})}$ order, and the Solid I- Solid II transition shares the spirit of vestigial transitions.
This points out that, from large to intermediate $V_2$, the increasing quantum fluctuations do not directly melt the $\rho_{\pm(\frac{\mathbf{b_1}}{4}+\frac{\mathbf{b_2}}{2})}$ order, but lead to a vestigial
$\rho_{\frac{\mathbf{b_1}}{2}}$ order.
This intriguing $T=0$ vestigial transition has not been reported in previous works~\cite{DNSheng2011boson,Luo2020boson} since they did not focus on the properties of states in the phase diagrams out of FCI, and missed these defining features of the Solid-I and Solid-II CDWs.
The complementary DMRG and ED results are shown in Appendix~\ref{appendix_b}.

\begin{figure}[htp!]
	\centering	
	\includegraphics[width=0.5\textwidth]{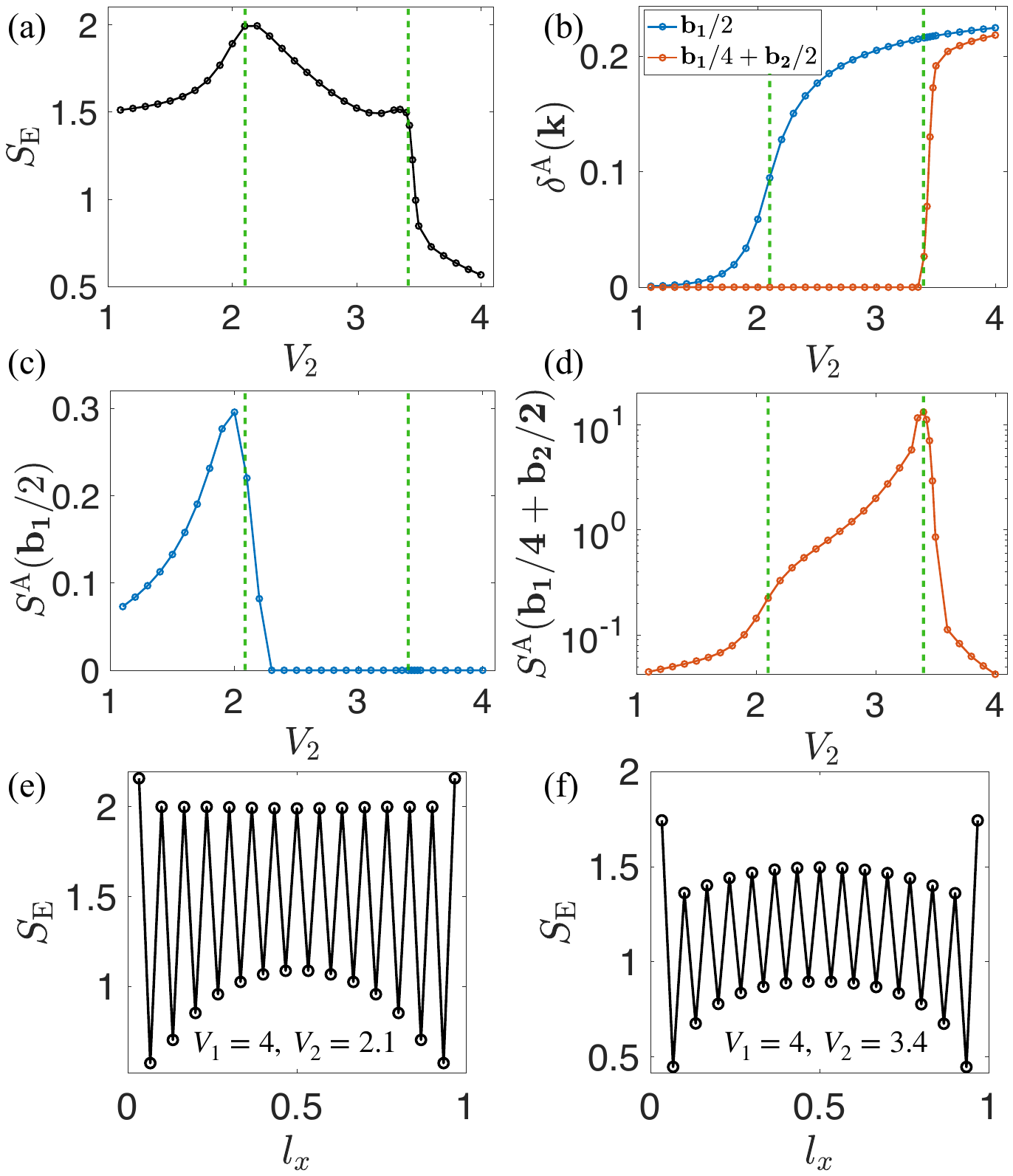}
	\caption{\textbf{The FCI-Solid I-Solid II sequence of transitions.} DMRG results on $N_y=4$ cylinders of (a) entanglement entropy $S_\mathrm{E}$, (b) Fourier transformation of real-space boson distribution $\delta^\mathrm{A}(\mathbf{k})$ at different momenta.
	The structure factors of boson density correlations at (c) $\mathbf{b_1}/2$ and (d) $\mathbf{b_1}/4+\mathbf{b_2}/2$	(in a log scale) are shown as well. 
	The green dashed lines are the two phase boundaries as in Fig.\ref{fig_fig7} (k).
	The entanglement entropies at the two transition points are shown in (e) and (f) respectively. 
	}
	\label{fig_fig8}
\end{figure}

We further demonstrate the FCI-Solid I-Solid II sequence of transitions in Fig.~\ref{fig_fig8}, and the behavior of entanglement entropies in Fig.~\ref{fig_fig8} (a) is in agreement with our phase diagram for the two-step transitions. 
More importantly, we will illustrate that the Solid I and Solid II states arise from the sequential condensation of multiple roton modes.
From the structure factor in Fig.~\ref{fig_fig7} (d), we can clearly observe that the value of $S^\mathrm{A}(\mathbf{q})$ at $\mathbf{b_1}/2$ (which is the roton minimum of this FCI) is stronger than those at other momenta. When approaching the FCI-Solid I transition point, the $S^\mathrm{A}(\mathbf{b_1}/2)$ continues to increase, as shown in Fig.~\ref{fig_fig8} (c), indicating the softening of the roton minimum.
Finally, the roton instability leads to the translational symmetry breaking and the consequent CDW order with the same wave vector of $\mathbf{b_1}/2$. As shown in Fig.~\ref{fig_fig8} (b), the order parameter $\delta^\mathrm{A}(\mathbf{b_1}/2)$ of the $\rho_{\frac{\mathbf{b_1}}{2}}$ order gradually increases and the order establishes at $V_2\sim2.1$.
After the formation of the $\rho_{\frac{\mathbf{b_1}}{2}}$ order, the fluctuations at $\mathbf{b_1}/2$ decreases towards 0.
As shown in the structure factor of Solid I ($\rho_{\frac{\mathbf{b_1}}{2}}$) [Fig.~\ref{fig_fig7} (e)], the structure factor (charge fluctuation) peaks at $\pm(\mathbf{b_1}/4+\mathbf{b_2}/2)$.
When further increasing $V_2$, the $\rho_{\frac{\mathbf{b_1}}{2}}$ order becomes stronger, accompanied by the increasing $S^\mathrm{A}(\pm(\mathbf{b_1}/4+\mathbf{b_2}/2))$, which manifests the softening of the $\pm(\mathbf{b_1}/4+\mathbf{b_2}/2)$ neutral modes in the Solid I state.
This increasing $\rho_{\pm(\frac{\mathbf{b_1}}{4}+\frac{\mathbf{b_2}}{2})}$ fluctuations finally leads to the Solid II state with $\rho_{\pm(\frac{\mathbf{b_1}}{4}+\frac{\mathbf{b_2}}{2})}$ order and the CDW fluctuations decrease towards 0 after the formation of $\rho_{\pm(\frac{\mathbf{b_1}}{4}+\frac{\mathbf{b_2}}{2})}$ order, as shown in Fig.~\ref{fig_fig8} (b,d) and Fig.~\ref{fig_fig7} (f).
To show the two transitions are triggered by the softening of neutral modes, we plot the entanglement entropies at the two transition points between insulators in Fig.~\ref{fig_fig8} (e,f), respectively. 
In contrast to the entanglement entropies of the three phases away from the transition points [Fig.~\ref{fig_fig7} (h-j)], the absent plateaus of $S_\mathrm{E}$ in the bulk of cylinders clearly suggest, at least the much reduced gaps. 
From the current results, the gap is more likely already closed at the Solid I-Solid II transition point as the both the upper and lower contours of the entanglement entropies support the gapless-like behaviors, compared to the FCI-Solid I transition point (the upper contour still supports a plateau).
Although whether these transitions are continuous or weakly first order might need more systematic simulations and analyses, the current results already strongly support that both of these transitions are induced by the softening of neutral modes.

More intriguingly, as shown in Fig.~\ref{fig_fig7} (d) and Fig.~\ref{fig_fig8} (d), the $\pm(\mathbf{b_1}/4+\mathbf{b_2}/2)$ fluctuations already appear as high-energy roton modes in the FCI state (although the lowest roton gap is at $\mathbf{b_1}/2$). 
The $S^\mathrm{A}(\pm(\mathbf{b_1}/4+\mathbf{b_2}/2))$ is also increasing during the FCI-Solid I transition and such softening modes become the lowest collective excitations in the Solid I state after the condensation of the condensation of $\mathbf{b_1}/2$ mode, until they finally condense as well and establish the $\rho_{\pm(\frac{\mathbf{b_1}}{4}+\frac{\mathbf{b_2}}{2})}$ order (we provide a complementary thermodynamic perspective in Fig.\ref{fig_figs5} of the Appendix to further support the $(\mathbf{b_1}/4+\mathbf{b_2}/2)$ fluctuations belong to higher-energy excitations than the $\mathbf{b_1}/2$ mode).
Compared with previous single-roton condensation~\cite{Mukherjee2022_rotoncondense, Kumar2022_condense,  LHY2024_roton_transition, LHY2024_FQAHS_transition} and the case in the checkerboard lattice in the earlier part of this work, to the best of our knowledge, this sequential multi-roton condensation in FCI is reported for the first time.

% ================================ %
\section{Discussions} \label{sec_discussions}
% ================================ %
The knowledge horizon of the interplay between FQH/FCI and symmetry-breaking states is expanding at an astonishing speed~\cite{GMP1986,  Kumar2022_condense, LHY2024_roton_transition, LHY2024_FQAHS_transition,Barkeshli2012CFL_FL, Barkeshli2014FQH-SF,Barkeshli2015FCI-SF, Song2023deconfined, songPhase2024}. This work is certainly among such a trend, and we here provide two exotic sequences of transitions out of FCI towards novel symmetry-breaking phases. 

In the case of $\nu=1/2$ FCI on checkerboard lattice, we find a sequence of FCI-GBDW-SS transitions, initially driven by the softening of the roton mode in FCI.
Intriguingly, the intermediate GBDW state with broken translational symmetry is gapless but not SF. 
Only when the CDW order in the GBDW state becomes strong enough, the further boson condensation at finite-momentum is triggered, leading to the LO-type SS with the same CDW order as that of GBDW.
More interestingly, as the bosons condense at $\pm(\pi/2,\pi/2)$ and the CDW wave vector of both SS and GBDW states is at ($\pi,\pi$), we find the CDW order is a secondary product order of the LO-type SF order, with the relationship $\rho_{(\pi,\pi)}\propto b^\ast_{(\frac{\pi}{2},\frac{\pi}{2})}b_{-(\frac{\pi}{2},\frac{\pi}{2})}$.
Therefore, the GBDW-SS transition is indeed a vestigial transition, as the increasing quantum fluctuations (from regions with strong to intermediate interactions) only partially melt the SF order while its vestigial CDW order survives with decreasing magnitude. 
This vestigial transition from the finite-momentum SS is similar in spirit to the physics of pair density waves where the thermal fluctuations often lead to vestigial CDW, nematicity, and charge-4e superconducting orders~\cite{Berg2009pdw, Fradkin2015interwined, Wang2020PDW}.
We also notice the effects of quenched disorder in pair density waves are studied, reporting the melted new glass states with charge-4e order~\cite{Mross2015pdwglass}.
Moreover, in the studies of superconductors with multiple broken symmetries, when thermal fluctuations melt the coherence of Cooper pairs, other symmetry-breaking orders and the Cooper pairs could still exist in the intermediate-temperature bosonic metal phase~\cite{Babaev2004metal,Bojesen2013metal,Bojesen2014metal,Babaev2015thermal,Babaev2021metal}.
Consequently, it would be meaningful for future work to study the full finite-temperature phase diagram of the GBDW-SS transition and see whether the gapless bosonic phase with broken translational symmetry could exist at the intermediate temperature of the finite-momentum SS and adiabatically connected to the GBDW phase in this $T=0$ ground-state phase diagram.
Furthermore, considering the realization of this unconventional SS, it might be hopeful to find the melted pair superfluid, similar to the charge-4e superconductors from pair density waves~\cite{Wu2023charge4e}.

We would like to further remark on this GBDW state, since to the best of our knowledge, most previous studies of bosonic non-SF states report no CDW orders~\cite{Paramekanti2002metal,Sheng2009metal, Block2011metal, Mishmash2011metal, Jiang2013metal,cao2024metal, cao2024metal2}.
However, this GBDW in our work is still compatible with previous scenarios for gapless non-SF states, such as the Bose metals that are often regarded as possible intermediate phases between superconductors and Mott insulators when the Cooper pairs lack phase coherence~\cite{Philip2003bosemetal, Philip2019bosemetal,Tsen2016metal, Tamir2019metal, Yang2019BoseMetal}. Since those superconductors do not have coexisting CDW orders, we would expect the emergence of similar translation-symmetry-breaking bosonic metals as intermediate phases in ground-state transitions related to superconductors (such as pair density waves)  with coexisting CDW orders.

The discovery of the GBDW state (gapless and non-SF state regardless of the CDW order), to the best of our knowledge, has not been reported in flat-band models, 
although previous attempts have been made to study possible Bose metals in moat-band models~\cite{sur2019metallic} or geometrically frustrated models~\cite{Hegg2021metal} which avoid single energy minima for free bosons, 
and we note the finite-momentum SS is also unconventional since it is driven by interaction in a flat-band model as well as the fact that the CDW order can be regarded as a secondary product of the SF order. 
These two exotic phases -- GBDW and finite-momentum SS -- discovered in this work suggest that the flat-band models, dominated by interactions instead of single-particle dispersions, are great synthetic platforms for studying exotic bosonic states.

In the case of $\nu=1/2$ FCI on honeycomb lattice, to the best of our knowledge, the sequential softening of the multiple roton modes is reported for the first time.
Moreover, the translation periods in Solid II along both momentum-space vectors are exactly twice of those in Solid I, and thus contributing to another fresh scenario of ground-state vestigial transitions.
The sequential FCI-Solid I-Solid II transitions, together with the FCI-GBDW-SS sequence of transitions, neither existing in literature, broaden the perspectives of the roton condensation in FCI, and we expect such scenarios could be discovered in future experiments as well, considering the fast-developing experiments on bosonic FQH/FCI~\cite{Julian2023photonFQH, Wang2024photonFQH}.  For example, it would be interesting for both theoretical and experimental explorations of the consequence of tuning neighboring interactions in the bosonic FCIs realized in square-lattice Hofstadter models and compare with our discoveries in the checkerboard-lattice model here.
On the other hand, it would be interesting to explore the deeper reasons that result in the different roton-driven transitions (gapless GBDW and SS in checkerboard lattice versus no gapless state found in the honeycomb lattice) in these two flat-band models.

As a motivation to discover the mechanism of transitions between FCI and SS, the other two scenarios mentioned in the introduction but still lacking enough evidence in literature, are also interesting for exploration.
We not only expect future discovery of a direct FCI-SS transition, but we would also like to remark on the second intuitive scenario, the FCI-SF-SS sequence of transitions.
Since the continuous FCI-SF transition has been achieved by tuning the band dispersion from the FCI at flat-band limit~\cite{LHY2024FCI_SF}, it would be interesting for future work to study whether the roton mode in FCI still exists in SF. 
It is possible that the softening of such roton modes (triggered by interaction, for example) in the SF phase will give rise to the CDW for the SS, and in this way, by tuning the band dispersion away from flatness might help to realize other different transitions from FCI towards SS with bosons possibly condensed at the single-particle energy minima of the dispersive bands and coexisting with CDW orders.
In fact, we have noticed the experimental preparation of SS states by softening of the roton mode in SF states, although not related to FCI yet~\cite{Chomaz2018roton, Tanzi2019supersolid, Fabien2019supersolid,Chomaz2019supersolid, zhang2019supersolid, Alana2023supersolid}.

% ================================ %

% =============================== %

% ================================ %
\section{Method}\label{sec_method}
In this work, we mainly use DMRG simulations~\cite{White1992_DMRG}, and we consider cylinders with $N_y\times N_x$ unit cells and total lattice sites $N=N_y\times N_x\times2$ for both lattices. We focus on the half filling of the lowest Chern band $\nu=N_\mathrm{b}/(N_y\times N_x)=1/2$ where $N_\mathrm{b}$ is the number of bosons. 
The main results are based on cylinders of $N_y=4$ unit-cell circumference and we consider up to $N_y=6$ for the case in checkerboard lattice. For the other direction with open boundary conditions, we consider up to $N_x=24$ on checkerboard lattice and $N_x=30$ on honeycomb lattice.
The DMRG simulations are based on the QSpace library~\cite{AW2012_QSpace} with U(1) symmetry for the conserved number of bosons. 
We keep the bond dimensions up to $D=8192$, and the maximum truncation error for $N_y=6$ is around $10^{-7}$.
We use XTRG~\cite{Chen2018XTRG} for thermodynamic simulations of $N_y=4$ cylinders, and the bond dimension is up to $D=1200$ with the maximum truncation error around $10^{-5}$.
% ================================ %

% ================================ %
\begin{acknowledgments}
{\it Acknowledgments}\,---\, We thank Kai Sun, Meng Cheng, and D. N. Sheng for inspiring discussions. HYL thank Shou-Shu Gong and Tian-Sheng Zeng for helpful discussions. HYL, BBC and ZYM acknowledge the support from the Research Grants Council (RGC) of Hong Kong Special Administrative Region of China (Project Nos. 17301721, AoE/P-701/20, 17309822, HKU C7037-22GF, 17302223), the ANR/RGC Joint Research Scheme sponsored by RGC of Hong Kong and French National Research Agency (Project No. A\_HKU703/22),  the GD-NSF (No. 2022A1515011007) and the HKU Seed Funding for Strategic Interdisciplinary Research. We thank HPC2021 system under the Information Technology Services and the Blackbody HPC system at the Department of Physics, University of Hong Kong, as well as the Beijng PARATERA Tech CO.,Ltd. (URL: https://cloud.paratera.com) for providing HPC resources that have contributed to the research results reported within this paper. HQW acknowledges the support from GuangDong Basic and Applied Basic Research Foundation (No. 2023B1515120013) and Youth S$\&$T Talent Support Programme of Guangdong Provincial Association for Science and Technology (GDSTA) (No. SKXRC202404). The ED calculations reported were performed on resources provided by the Guangdong Provincial Key Laboratory of Magnetoelectric Physics and Devices, No. 2022B1212010008.
\end{acknowledgments}

\bibliographystyle{apsrev4-2}
%

%\iffalse
% ================================================================================
% appendix
% ================================================================================
\clearpage

\renewcommand{\theequation}{S\arabic{equation}} \renewcommand{\thefigure}{S%
\arabic{figure}} \setcounter{equation}{0} \setcounter{figure}{0}

\appendix

\section{Supplementary Information of the checkerboard lattice model} \label{appendix_a}
In the main text, we have mentioned that the limit bond dimenions in DMRG simulations might lead to inaccurate CDW orders in both the GBDW and the SS phases.
Here, we take the GBDW phase at $V_1=1$ and $V_2=1.5$ as an example.

\begin{figure}[htp!]
	\centering	
	\includegraphics[width=0.32\textwidth]{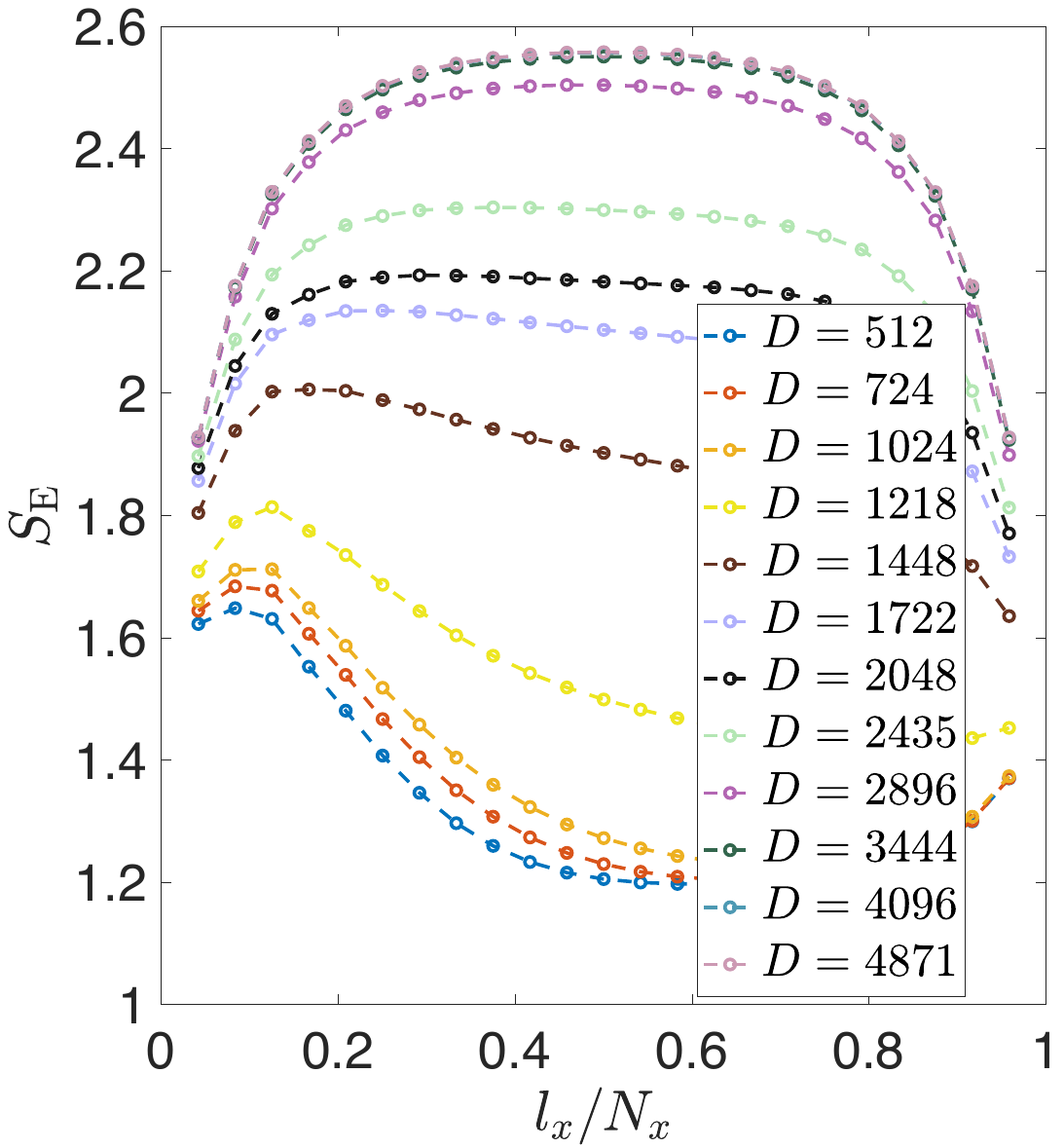}
	\caption{\textbf{Entanglement entropy of GBDW with different bond dimensions.} This is the result of a $4\times24\times2$ cylinder with $V_1=1$ and $V_2=1.5$, and the entanglement entropy is not able to converges until $D>4000$. 
	}
	\label{fig_figs1}
\end{figure}

\begin{figure}[htp!]
	\centering	
	\includegraphics[width=0.5\textwidth]{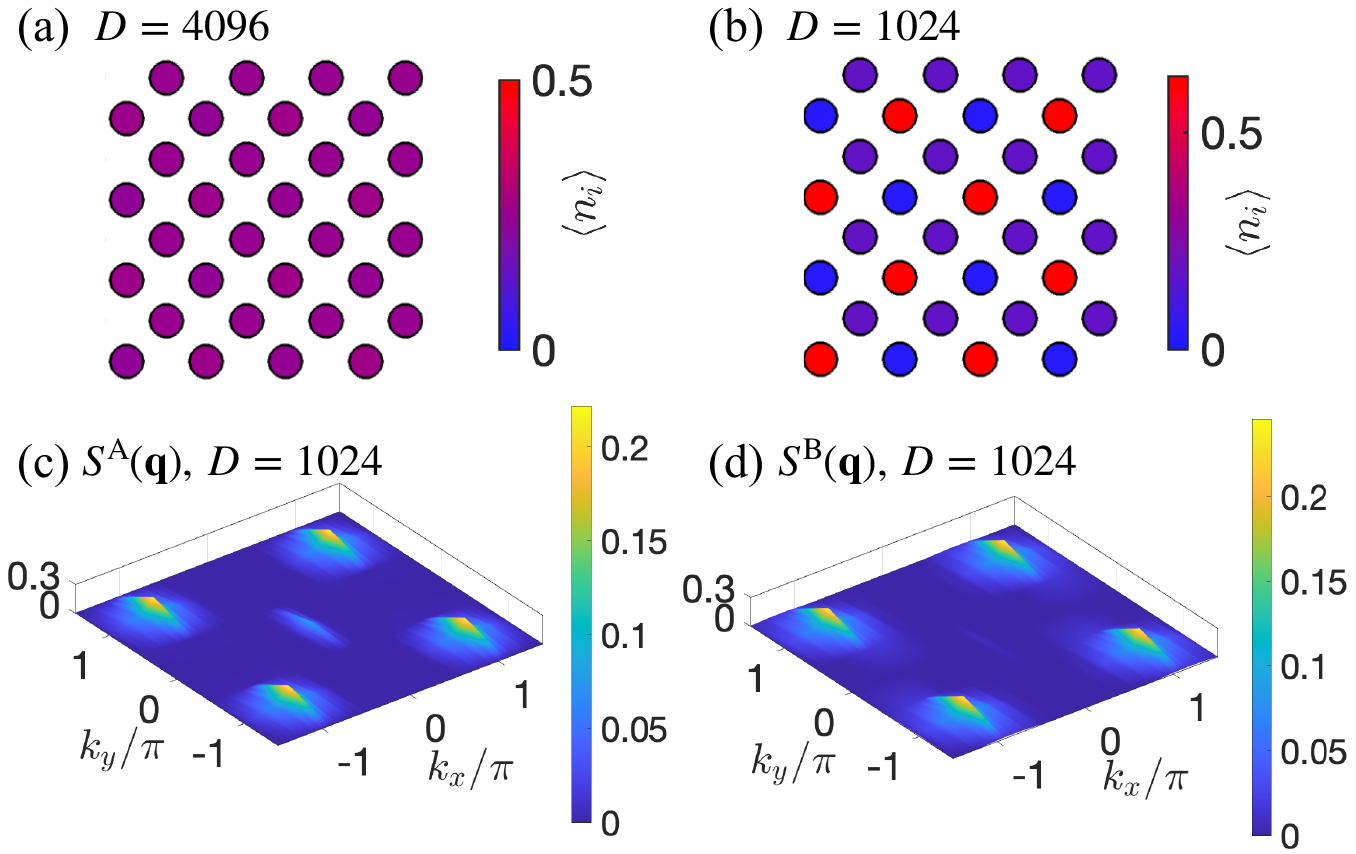}
	\caption{\textbf{Boson distributions and structure factors of GBDW with different bond dimensions.} This is the result of a $4\times24\times2$ cylinder with $V_1=1$ and $V_2=1.5$. The real-space boson distributions are shown for (a) $D=4096$ and (b) $D=1024$. The structure factors from both sublattices at $D=1024$ are shown in (c,d) respectively.
	}
	\label{fig_figs2}
\end{figure}

As shown in Fig.~\ref{fig_figs1}, the entanglement entropy will only converge at large enough bond dimensions, and when the bond dimension is too small, other physical observables are not converged either or even far from compatible with the converged results.
The real-space boson occupation numbers at different bond dimensions are shown in Fig.~\ref{fig_figs2} (a,b) respectively. It is well converged when $D=4096$ for this $4\times24\times2$ cylinder, and the results are the same in the main text that the real-space distribution is uniform and the long-range CDW order is manifested from the structure factors [Fig.~\ref{fig_fig2} (c,d) and Fig.~\ref{fig_fig3} (c)].
However, when the bond dimension is too small, the CDW order parameter would be incorrect, as shown in Fig.~\ref{fig_figs2} (b), where one sublattice is disordered while the other one has a $\rho_{(\pi,\pi)}$ order. 
Besides, the total numbers of bosons in two sublattices are not equal when $D=1024$, and the average number of bosons per site in the disordered sublattice is much smaller than $\nu/2$, which is the same as the result in the previous iDMRG work~\cite{Luo2020boson}.
The structure factors of the both sublattices when $D=1024$ are shown in Fig.~\ref{fig_figs2} (c,d), and the values at ($\pi,\pi$) are much smaller than the converged values [Fig.~\ref{fig_fig3} (c)], further clarifying that the limited bond dimension would be hard to capture the exact $\rho_{(\pi,\pi)}$ orders of both sublattices or the strong density-density fluctuations in the GBDW state.

\section{Supplementary Information of the honeycomb lattice model} \label{appendix_b}

\begin{figure}[htp!]
	\centering	
	\includegraphics[width=0.5\textwidth]{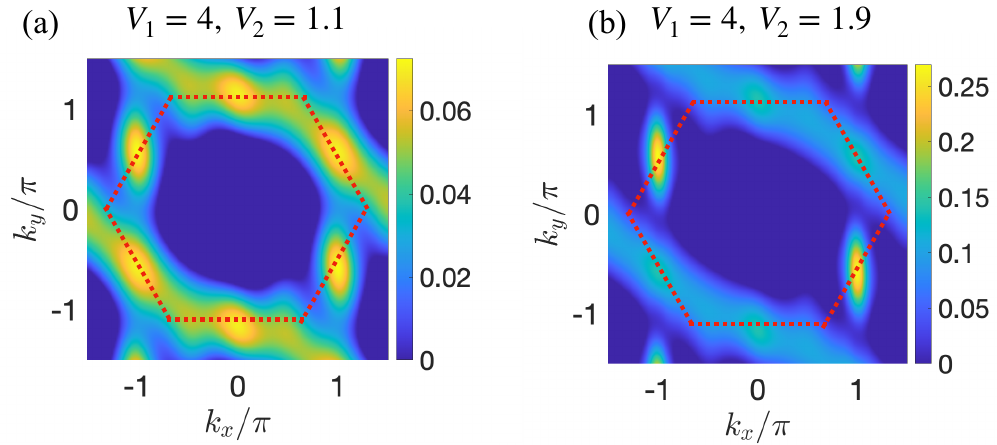}
	\caption{\textbf{Supplementary structure factors $S^\mathrm{A}(\mathbf{q})$ of FCI} at (a) $V_1=4,\ V_2=1.1$ and (b) $V_1=4,\ V_2=1.9$. 
	}
	\label{fig_figs3}
\end{figure}

In Sec.~\ref{sec_HC}, we have shown that the CDW order of Solid I is triggered from the roton instability at $\mathbf{b_1}/2$. 
And some complementary results of structure factors of boson density correlations are presented in Fig.~\ref{fig_figs3}.  When it is far from the FCI-Solid I transition point ($V_1=4,\ V_2=1.1$), it is clear that the value of $S^\mathrm{A}(\mathbf{q})$ at $\mathbf{b_1}/2$ is almost the same as those at other $M$ points of the Brillouin zone.
When approaching the FCI- Solid II transition point ($V_1=4,\ V_2=1.9$), the DMRG simulations would pick the translation symmetry breaking along the $\vec{a}_1$ direction, so we see the value of $S^\mathrm{A}(\mathbf{q})$ at $\mathbf{b_1}/2$ is getting much larger than the others, and its instability finally leads to the $\rho_{\frac{\mathbf{b_1}}{2}}$ order.

\begin{figure}[htp!]
	\centering	
	\includegraphics[width=0.4\textwidth]{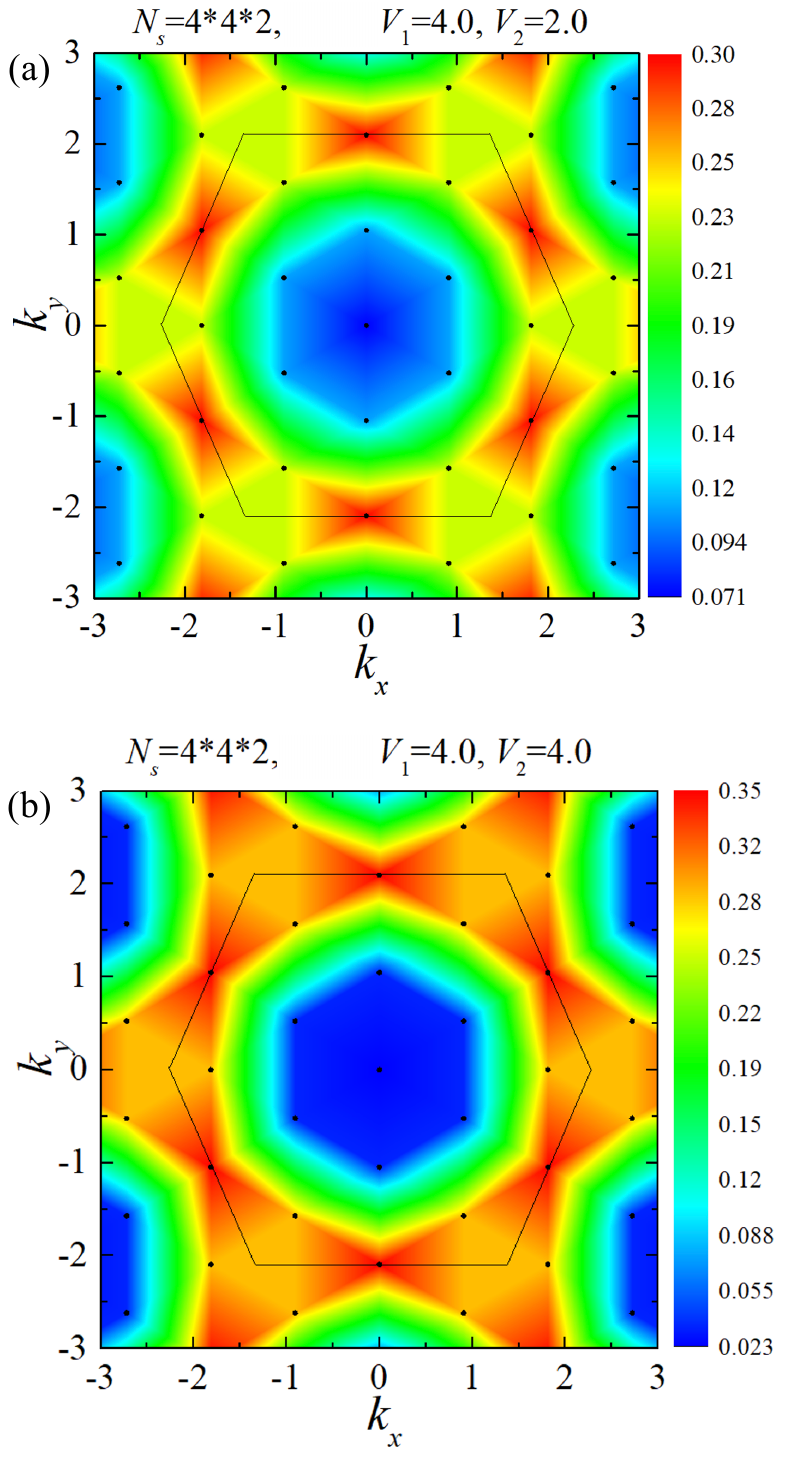}
	\caption{\textbf{ED results of structure factors $S^\mathrm{A}(\mathbf{q})$} at (a) $V_1=4,\ V_2=2$ and (b) $V_1=4,\ V_2=4$. 
		The black dots refer to the considered momenta in the ED simulations of this $4\times4\times2$ torus.
	}
	\label{fig_figs4}
\end{figure}

\begin{figure}[htp!]
	\centering	
	\includegraphics[width=0.4\textwidth]{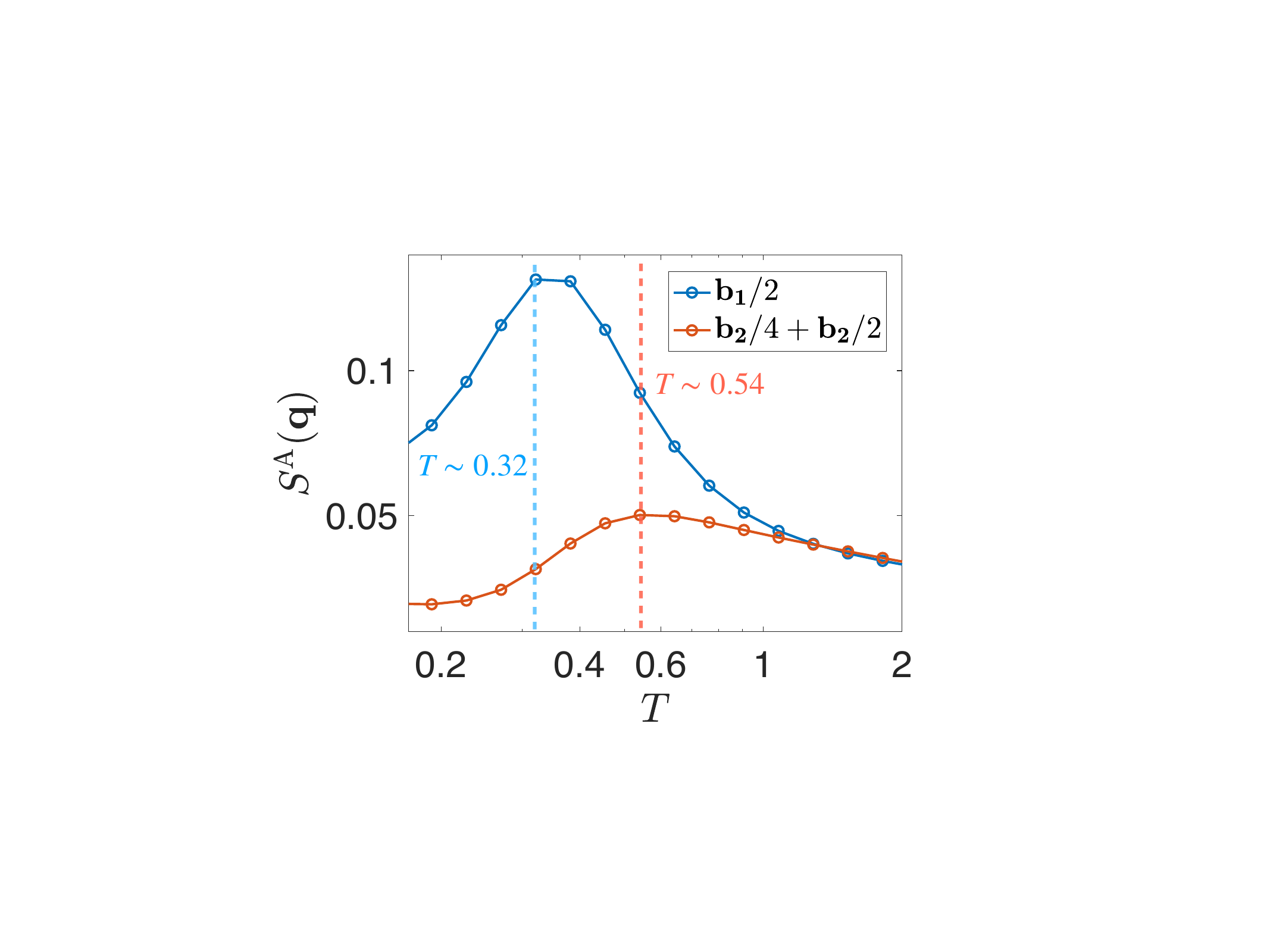}
	\caption{\textbf{XTRG results of structure factors as functions of temperature.} The result is from a $4\times12\times2$ cylinder and the peaks of the two curves are at different temperatures, indicating the neutral fluctuations of the $(\mathbf{b_1}/4+\mathbf{b_2}/2)$ mode belong to higher-energy excitations than the $\mathbf{b_1}/2$ mode.
	}
	\label{fig_figs5}
\end{figure}

To demonstrate the CDW degeneracy would be preserved in the Solid I and the Solid II states if the geometry is symmetric, we show complementary ED results of a $4\times4\times2$ torus.
The intra-sublattice structure factors in Fig.~\ref{fig_figs4} qualitatively agree with our DMRG simulations and conclusions (while overlooked in previous works), although the transition points determined in previous ED work from the ground-state fidelity susceptibility are different.
When $V_2$ is small, the CDW order is just manifested from the peaks at $\mathbf{b_1}/2$ and the other $M$ points with $C_3$ rotations. Only when $V_2$ is getting larger, the translation periods along both directions further double, and thus the structure factors at $\pm(\mathbf{b_1}/4+\mathbf{b_2}/2)$ are almost the same as $\mathbf{b_1}/2$.

Furthermore, we have shown that the sequence of FCI-Solid I-Solid II transitions is from the progressively softening of neutral modes in FCI from ground-state results. 
Here, we provide an additional thermodynamic perspective to support our conclusion in Fig.\ref{fig_figs5}. We show the temperature-dependent structure factors of the two relevant modes, which clearly show that the $(\mathbf{b_1}/4+\mathbf{b_2}/2)$ mode belongs to higher-energy excitations than the $\mathbf{b_1}/2$ mode.
%\fi

\end{document}